\documentclass[11pt]{article}
\usepackage[a4paper,margin=1in]{geometry}
\usepackage[T1]{fontenc}
\usepackage{lmodern}
\usepackage[utf8]{inputenc}
\usepackage{graphicx}
\usepackage{microtype}
\usepackage{booktabs}
\usepackage{array}
\usepackage{amsmath,amssymb}
\usepackage{xcolor}
\usepackage{hyperref}
\definecolor{linkblue}{RGB}{0,70,140}
\definecolor{citered}{RGB}{170,0,0}
\hypersetup{colorlinks=true,linkcolor=black,citecolor=citered,urlcolor=linkblue}
\usepackage{caption}
\usepackage{subcaption}
\usepackage{tabularx}
\usepackage{setspace}
\usepackage{authblk}
\usepackage{siunitx}
\usepackage[numbers,sort&compress]{natbib}
\graphicspath{{figures/}}

\newcommand{\MeanControlContributionVone}{3.88}
\newcommand{\MeanCostZeroContributionVone}{6.99}
\newcommand{\MeanCostOneContributionVone}{6.44}
\newcommand{\MeanCostThreeContributionVone}{5.12}

\newcommand{\NIncludedParticipantsVone}{380}
\newcommand{\NIncludedGroupsVone}{76}

\newcommand{\NDemoMatchedVone}{379}

\newcommand{\NDemoTotalVone}{380}
\newcommand{\DemoAgeMeanVone}{30.58}
\newcommand{\DemoAgeSdVone}{9.79}

\newcommand{\NormControlGroupMeanVone}{4.76}

\newcommand{\NormCostZeroGroupMeanVone}{7.59}

\newcommand{\NormCostOneGroupMeanVone}{6.77}

\newcommand{\NormCostThreeGroupMeanVone}{5.87}

\newcommand{\NormGroupKWPVone}{0.00162}

\newcommand{\NormDemocraticSpearmanRhoVone}{-0.31}
\newcommand{\NormDemocraticSpearmanPVone}{0.01801}

\newcommand{\DeltaOwnMeanControlVone}{0.561}

\newcommand{\DeltaOwnMeanCostZeroVone}{0.061}

\newcommand{\DeltaOwnMeanCostOneVone}{0.425}

\newcommand{\DeltaOwnMeanCostThreeVone}{0.500}

\newcommand{\DeltaOthersMeanControlVone}{0.522}

\newcommand{\DeltaOthersMeanCostZeroVone}{-0.175}

\newcommand{\DeltaOthersMeanCostOneVone}{0.189}

\newcommand{\DeltaOthersMeanCostThreeVone}{0.272}

\newcommand{\GammaMeanControlVone}{0.846}

\newcommand{\GammaMeanCostZeroVone}{0.606}

\newcommand{\GammaMeanCostOneVone}{0.210}

\newcommand{\GammaMeanCostThreeVone}{0.714}

\newcommand{\DeltaSelfMinusOthersMeanControlVone}{0.038}

\newcommand{\DeltaSelfMinusOthersMeanCostZeroVone}{0.236}

\newcommand{\DeltaSelfMinusOthersMeanCostOneVone}{0.236}

\newcommand{\DeltaSelfMinusOthersMeanCostThreeVone}{0.228}

\newcommand{\DeltaVCMeanCostZeroVone}{-0.511}

\newcommand{\DeltaVCMeanCostOneVone}{0.039}

\newcommand{\DeltaVCMeanCostThreeVone}{1.245}

\newcommand{\DeltaVDMeanCostZeroVone}{0.078}

\newcommand{\DeltaVDMeanCostOneVone}{0.350}

\newcommand{\DeltaVDMeanCostThreeVone}{0.649}

\newcommand{\DeltaVAllMeanCostZeroVone}{-0.289}

\newcommand{\DeltaVAllMeanCostOneVone}{0.184}

\newcommand{\DeltaVAllMeanCostThreeVone}{-0.053}

\newcommand{\DeltaVNMeanCostZeroVone}{0.289}

\newcommand{\DeltaVNMeanCostOneVone}{-0.301}

\newcommand{\DeltaVNMeanCostThreeVone}{-1.394}

\title{Participation Costs Narrow Democratic Cooperation}
\author[1,2,3]{Mohammad Salahshour\thanks{\texttt{msalahshour@ab.mpg.de}}}
\author[1,2,3]{Fjolle Shabani}
\author[2,4,5,6]{Urs Fischbacher}
\author[1,2,3]{Iain D. Couzin}
\affil[1]{Department of Collective Behaviour, Max Planck Institute of Animal Behavior, Konstanz, Germany}
\affil[2]{Centre for the Advanced Study of Collective Behaviour, University of Konstanz, Konstanz, Germany}
\affil[3]{Department of Biology, University of Konstanz, Konstanz, Germany}
\affil[4]{Department of Economics, University of Konstanz, Konstanz, Germany}
\affil[5]{Thurgau Institute of Economics, Kreuzlingen, Switzerland}
\affil[6]{CESifo, Munich, Germany}
\date{}

\begin{document}
\maketitle

\begin{abstract}

Collective action often requires institutions that make cooperation individually worthwhile. We ask whether democratic allocation of public-good return can transform a repeated public good into a self-sustaining cooperative institution, and how participation costs reshape that process. A simple evolutionary model shows that voted redistribution can support a prosocial allocation order, but can also sustain an antisocial allocation order or democratic free riding, in which individuals benefit from an institution maintained by others while avoiding the cost of participation. The model predicts competing effects of voting cost. Cost can suppress use of the institution to reward low contributors under strong selection, but can also thin the active electorate and erode contributor-rewarding support. We test these predictions in a preregistered online experiment with \NIncludedGroupsVone{} five-person groups. Endogenous democratic redistribution increased contributions relative to an equal-share public-goods control, with zero-cost voting producing the strongest temporal improvement. Voting costs did not mainly turn active voters toward low-contributor-rewarding allocation. Instead, they shifted behavior toward abstention and democratic free riding, made abstention locally rewarding, and widened the gap between post-task perceptions of democratic participation and the behavioral record. Democratic allocation can therefore stabilize cooperation, but participation costs can reduce the number of people actively sustaining the institution and can make that erosion less visible to participants themselves.
\end{abstract}

\section{Introduction}
Collective action problems arise when group welfare depends on costly individual effort but the benefits can still be shared with non-contributors \cite{Hardin,Olson,Ostrom,Dietz}. Successful solutions to these problems underlie many social institutions, from common-pool governance to public-goods provision, yet they remain vulnerable to free riding. Institutions can play a constructive role by changing incentives, coordinating expectations, and making cooperation more durable \cite{Ostrom,DannenbergGallier}. However, institutional success depends on more than formal design because it also depends on who keeps using the institution, whether participants believe others are using it in good faith, and whether continued participation still feels worth the effort.

Much previous work on institutional solutions has focused on sanctioning and related mechanisms that make free riding costly \cite{Fehr,SalahshourPunishment2021,SalahshourCostlyInstitutions2021,SutterInstitution,DannenbergGallier}. Democracy offers a different route. Instead of imposing incentives from outside the group, democratic allocation lets group members decide how the collectively produced good should be distributed. Such procedures can increase legitimacy and cooperation because people help choose the rules that govern them \cite{DalBoFosterPutterman,KameiDemocracy,Olken2010,GrossmanBaldassarri}. Yet democracy is not just a rule for allocation; it is also an activity. Voting, deliberating, monitoring, and coordinating require time, attention, money, or effort. Once participation is costly, democracy can create a second dilemma in which people benefit from a cooperative rule that others maintain while avoiding the cost of taking part themselves \cite{Palfrey,Goeree,Krasa}. Unequal voice and unequal participation can therefore determine who governs and the direction in which the institution is heading \cite{Lijphart1997,Verba1995,MettlerSoss2004,SchlozmanVerbaBrady2012,Prior2007,KarpowitzMendelbergShaker2012}.

Recent democratic discontent is often visible as declining confidence in institutions, distrust and hostility between political groups, and disputes over whether collective decisions are legitimate \cite{Norris2011,FoaMounk2016,GrahamSvolik2020,Bermeo2016,LevitskyZiblatt2018,LuhrmannLindberg2019}. Underlying many such problems may not be a simple behavioral malfunction, but deeper gaps in how actors using democratic means perceive incentives, norms, and one another's roles, including who is cooperating, which participation norms should apply, whether others are exploiting the institution, and whether continued participation is worthwhile or legitimate. Public-goods experiments show that beliefs about others' behavior shape contribution, norm-elicitation studies show that behavior depends on what people think ought to be done, and biased perceptions can trap groups in inefficient collective states \cite{FischbacherGaechter2010,BicchieriXiao2009,KrupkaWeber2013,SantosLevinVasconcelos,OberhauserCouzinSalahshour2025}. Democratic cooperation therefore depends on three linked ingredients, with incentives making cooperation worthwhile, participation keeping the institution active, and perceptions informing participants what others are doing and what the institution represents.

We ask when these ingredients allow democratic allocation of public-good return to transform a public good into a self-sustaining cooperative institution, and when participation costs instead create democratic free riding. To answer this question, we combine a simple evolutionary model, an exploratory behavioral model, and a preregistered online public-goods experiment. The evolutionary model asks when democratic allocation of public-good return can turn a public good into a self-sustaining cooperative institution, and when participation cost instead creates democratic free riding. The experiment tests the corresponding behavioral predictions in human groups. Participants contributed to a common pool in groups of five over 25 rounds. In three democratic redistribution treatments, they chose whether to cast an active vote for redistribution to participants whose contribution was above or equal to the group average, redistribution to participants whose contribution was below or equal to the group average, equal redistribution among all group-members, or to abstain. Voting cost was experimentally manipulated at 0, 1, or 3 Money Units (MUs), alongside an equal-share public-goods control without voting \cite{Salahshour2024}. After the task, participants reported their own contribution, others' contribution, appropriate contribution, and, in the democratic treatments, their own use of each voting option.

Endogenous democratic redistribution increased cooperation relative to the equal-share public-goods control in all three voting treatments. Higher participation cost, however, changed the institution from within by reducing active democratic use, shifting dominant voting types from contributor-rewarding participation toward abstention and democratic free riding, and producing larger gaps between reported and recorded democratic participation. These results are consistent with the weaker-selection participation-thinning regime of the evolutionary model, and an exploratory behavioral model links the experimental pattern back to the same participation mechanism. Individual-difference and demographic measures added context, but they did not replace the institutional result, in which voting cost, contribution behavior, and perception were the main organizers of democratic use. Our results show that democratic redistribution can stabilize cooperation, yet participation cost narrows the democratic coalition that sustains it.

\section{Results}
\subsection{Theory of democratic allocation of public-good return and costly participation}
Can democratic allocation of public-good return turn a public good into a self-sustaining cooperative institution, and what happens when using that institution is itself costly? We first study this question in a binary evolutionary model. The model formalizes three possible uses of democratic allocation of public-good return, covering a prosocial allocation order that rewards contributors and supports cooperation, an antisocial allocation order that rewards low contributors and undermines cooperation, and democratic free riding, in which individuals benefit from a contributor-rewarding institution while avoiding the cost of participation.

\begin{figure}[p]
    \centering
\includegraphics[width=\linewidth]{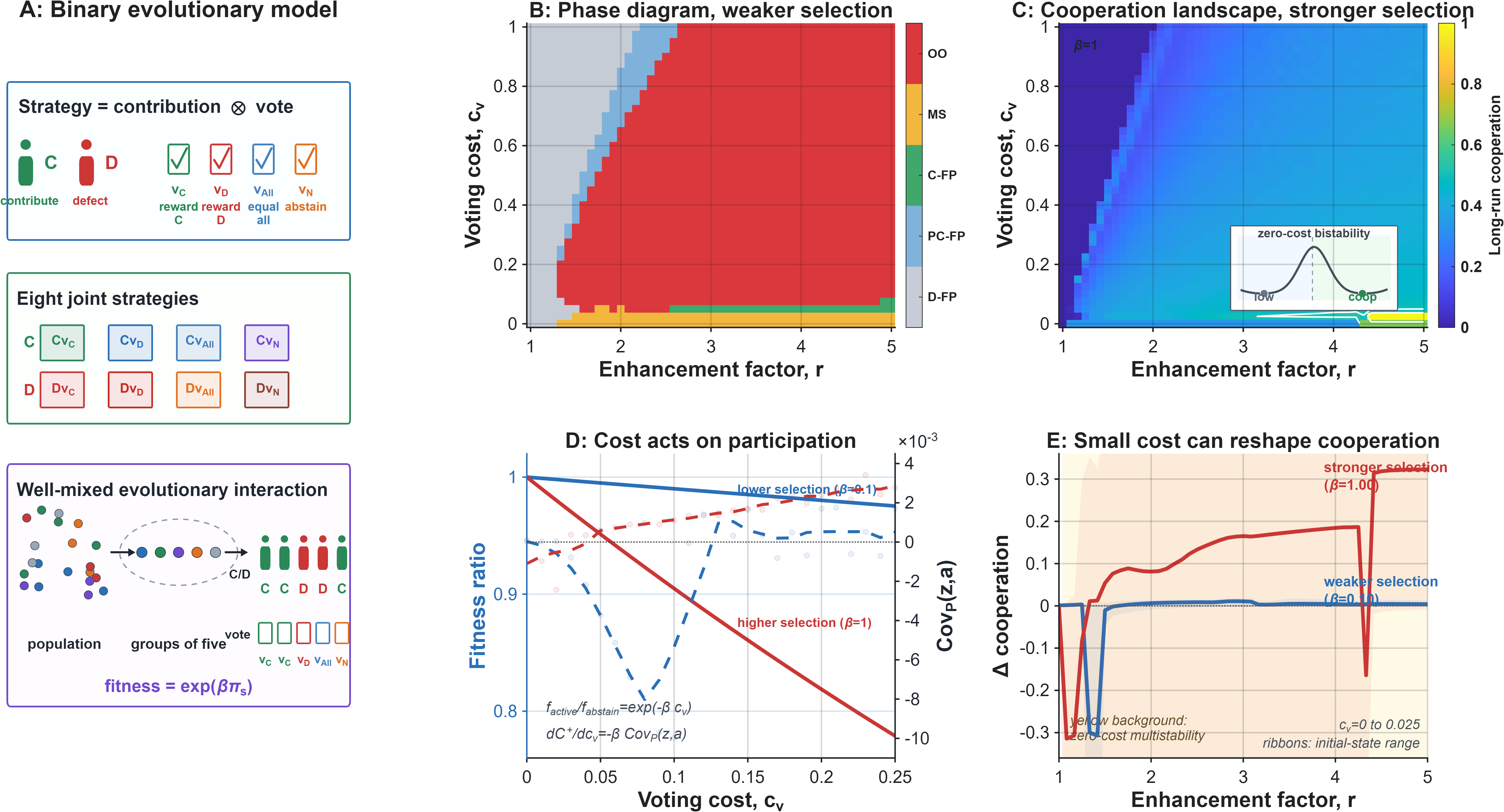}
\caption{Evolutionary model of costly democratic participation. \textbf{A:} Binary evolutionary model structure. Individuals combine a contribution strategy with a voting strategy, groups of five are sampled from a well-mixed population, and payoff is mapped to fitness. \textbf{B:} Replicator phase diagram for the binary evolutionary model in the weaker-selection regime. D-FP denotes a defective fixed point, where low-contribution strategies dominate. PC-FP denotes a partially cooperative fixed point, where cooperation persists at an intermediate level. C-FP denotes a cooperative fixed point, where contributor-rewarding cooperation is stable. MS denotes an initial-condition-dependent multistable regime, where different starting states reach different attractors. OO denotes an oscillatory-orbit regime, where stationary-window dynamics remain oscillatory. \textbf{C:} Long-run cooperation in the stronger-selection analytical model across the enhancement factor and voting-cost plane, averaged over specified initial states. The inset gives the schematic zero-cost bistability interpretation. \textbf{D:} Direct participation filter. Voting cost penalizes active voting relative to abstention with strength $f_{\mathrm{active}}/f_{\mathrm{abstain}}=\exp(-\beta c_v)$, and its immediate effect on cooperation is $-\beta\,\mathrm{Cov}_P(z,a)$. The right axis plots the smoothed stationary contribution--participation covariance across the $r=4$ voting-cost range, with faint points showing the underlying grid values. \textbf{E:} Small-cost cooperation response. Thick lines show the change in long-run cooperation from $c_v=0$ to $c_v=0.025$, averaged over the same initial states. Blue and red ribbons show the full range across those initial states for weaker and stronger selection, respectively. The pale yellow background marks $r$ values that are multistable at zero cost under stronger selection. Binary-model parameters were group size $g=5$, contribution cost $c=1$, investment and voting mutation rates $10^{-3}$, $\beta=0.10$ for weaker selection and $\beta=1.00$ for stronger selection, $T=6000$ evolutionary update steps with a stationary window of 1200 steps for panels B, C, and E, and $T=9000$ steps with a stationary window of 2500 steps for panel D.}
    \label{fig:binary-exponential-theory}
\end{figure}

In our model, individuals interact in groups of five and choose both an investment strategy and a voting strategy. They may contribute to the public good or defect; contributions are multiplied by an enhancement factor $r$ and the resulting public-good return is divided according to democratic allocation of public-good return. Individuals can vote to allocate this return to contributors, to defectors, equally among all, or abstain. This gives eight joint strategies (Fig.~\ref{fig:binary-exponential-theory}A). Payoff is mapped to fitness through $f_s=\exp(\beta\pi_s)$, where $\beta$ is the strength of selection. Weaker selection makes payoff differences bias strategy change gradually, whereas stronger selection amplifies locally advantageous strategies more sharply; see Methods for details.

We study the model using replicator dynamics and finite-population simulations (Supplementary Information (SI) Sections~S1.5 and S1.6). Even this minimal institution can settle into several collective regimes, including defective, partially cooperative, and cooperative fixed points, multistable regions, and persistent oscillations (Fig.~\ref{fig:binary-exponential-theory}B). Cooperative democracy emerges only when public-good incentives and active electoral composition jointly support a contributor-rewarding allocation rule. When participation is free, such a democracy can evolve, though not inevitably. At low voting cost, the same parameter region can be multistable, and dynamics can converge either to a flourishing cooperative democracy or to an institution that rewards low contributors (Fig.~\ref{fig:binary-exponential-theory}C). This raises the question of how institutional design regulates these possible outcomes. A natural conjecture is that a small cost of voting could improve free democracy by discouraging use of the institution to reward low contributors. Whether this filtering effect dominates in the evolutionary dynamics depends on the strength of selection and the composition of the active electorate.

A small positive cost can sometimes increase cooperation, but only under particular evolutionary conditions. When zero-cost democracy is dominated by low-contribution strategies that use active voting to reward low contributors, a small participation cost thins that use more than contributor-rewarding use, raises the contributor-rewarding share among the remaining active electorate, and can move the system toward higher cooperation (Fig.~\ref{fig:binary-exponential-theory}D,E). This can happen under stronger selection, where local payoff differences are amplified and the zero-cost electorate can be dominated by low-contributor-rewarding voters. Under weaker selection, the active electorate in free democracy is already more contributor-rewarding in composition. In that case, voting cost mainly removes participants from the cooperative electorate, increases abstention, and narrows contributor-rewarding support. The model therefore predicts a tension between two effects: cost can filter out active use that rewards low contributors when selection is sharp and the zero-cost electorate is poorly composed, but it can also hollow out the participation that sustains cooperation. See Methods and SI Sections~S1.3 and S1.4 for the derivation, and SI Section~S1.5 for the finite-population analysis.

\begin{figure}[p]
\centering
\includegraphics[width=\textwidth]{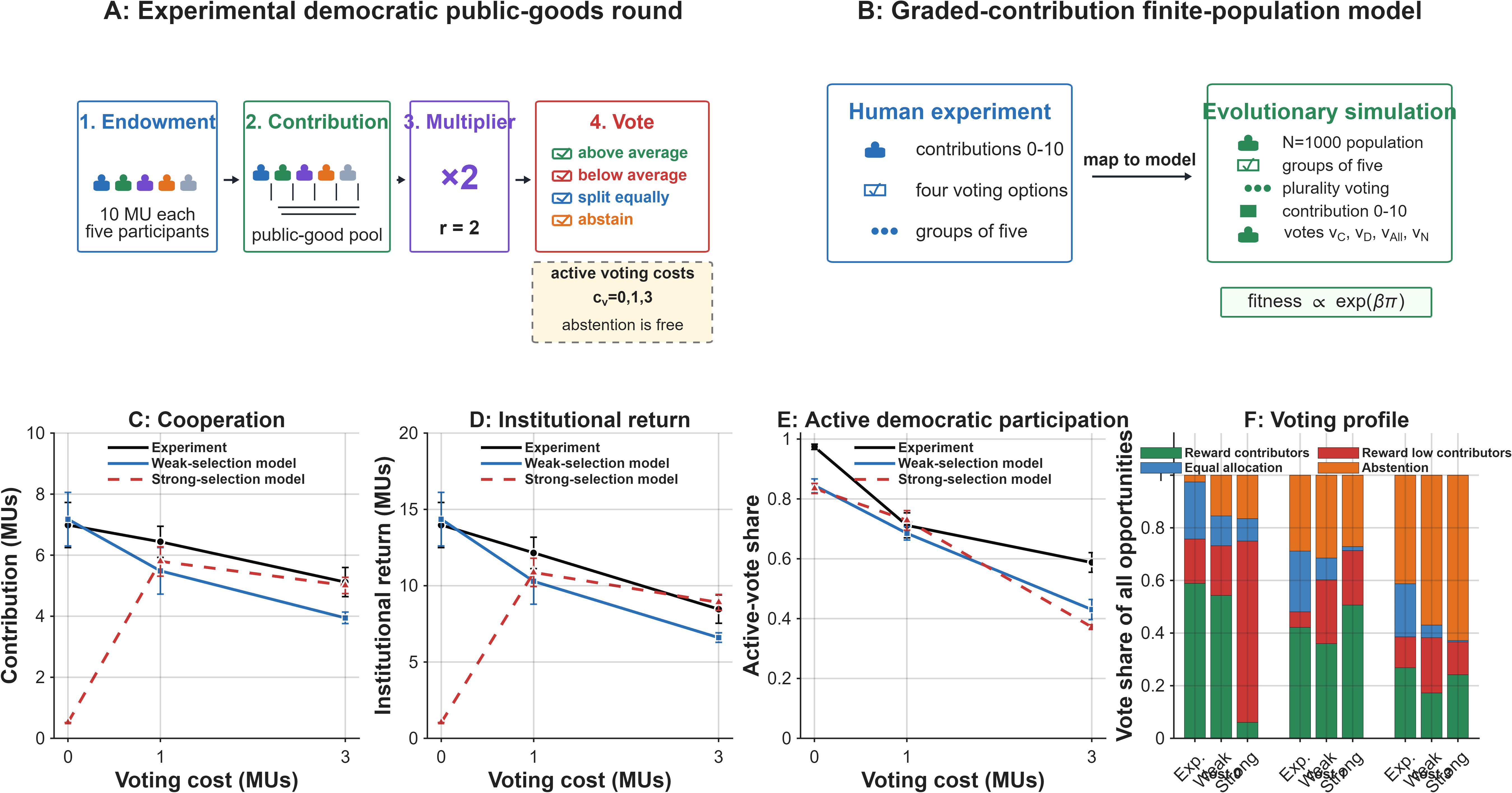}
\caption{Graded-contribution bridge from the evolutionary model to the human democratic public-goods task. \textbf{A:} Experimental round structure. Participants received an endowment, chose a 0--10 MU contribution, the public good was doubled, and democratic-treatment participants chose whether to vote for allocation to above-or-equal-to-average contributors, allocation to below-or-equal-to-average contributors, equal allocation among all group members, or abstain. Active voting cost was 0, 1, or 3 MUs; abstention was free. \textbf{B:} Mapping from the human task to the graded-contribution finite-population model. \textbf{C-F:} Human experiment compared with weaker- and stronger-selection stationary simulations for contribution (C), institutional return, defined as allocated public-good share minus voting cost (D), active voting (E), and voting composition, including active electorate composition and abstention (F). Simulation time averages are computed over the last 1000 of $T=5000$ generations. Simulations use well-mixed populations, groups of five, $N=1000$, endowment 10, multiplier $r=2$, and voting costs $c_v\in\{0,1,3\}$. The weaker-selection model uses $\beta=0.008$, contribution mutation $\mu_C=0.120$, vote mutation $\mu_V=0.003$, and initial vote probabilities $(0.70,0.10,0.15,0.05)$; the stronger-selection contrast sets $\beta=0.10$ and keeps the same mutation and initialization structure. Error bars show standard errors across experimental groups or simulated replicate populations.}
\label{fig:evolutionary-selection-comparison}
\end{figure}

The binary model has the analytical advantage of tractability, but many collective-action settings allow graded rather than all-or-none contribution. We therefore asked whether the same cost-dependent phenomena extend to a graded-contribution finite-population model. The model uses integer contributions from 0 to 10, well-mixed groups of five, plurality voting among three active allocation rules, abstention, random active-vote tie breaking, and a random all-abstention fallback (Fig.~\ref{fig:evolutionary-selection-comparison}A,B; Methods). In the weaker-selection graded-contribution regime, increasing voting cost reduces contribution, institutional return, active voting, and contributor-rewarding support, while increasing abstention (Fig.~\ref{fig:evolutionary-selection-comparison}C-F). Stronger selection more often produces an intermediate-cost cooperation peak because the zero-cost active electorate can be dominated by low-contribution strategies using the institution to reward low contributors. This makes the contribution--participation covariance negative, so a small voting cost initially penalizes the strategies maintaining the low-cooperation state. Initial-condition, population-size, and long-run time-series screens show the same separation (SI Section~S1.6). Monotone cost penalties are most robust under weaker selection, whereas positive-cost optima are more common under sharper selection and stronger initial-condition dependence.

These results raise the question of which regime governs democratic cooperation in human groups. If human behavior resembles the stronger-selection filtering regime, a modest voting cost should improve cooperation by suppressing low-cooperation active strategies. If it resembles the weaker-selection participation-thinning regime, voting cost should mainly reduce active participation, increase abstention, and gradually narrow democratic cooperation.


\subsection{Experimental design}
We tested these predictions in a preregistered online experiment \cite{Salahshour2024} on Prolific \cite{PalanSchitter2018} with four conditions, including a standard equal-share public-goods control without voting and three democratic redistribution versions of the same game in which voting cost 0, 1, or 3 MUs. These values were chosen to compare a frictionless democratic institution with modest and higher participation costs. The primary sample contained 18 groups in the no-democracy control, 18 in cost-0 democracy, 21 in cost-1 democracy, and 19 in cost-3 democracy. Participants played in groups of five for 25 rounds. In each round, each person received 10 MUs, chose how much to contribute to a common pool, and, in the democratic conditions, voted on how the multiplied public good should be redistributed. In the no-democracy control, the multiplied public good was divided equally among all five group-members. The procedure, incentives, and post-task survey sequence are reported in Methods and SI Sections~S3, S7, S10, and S9.

The Results are organized from confirmatory to exploratory evidence. We first report the preregistered cooperation, institutional returns, participation, and dynamic tests. We then use the exploratory behavioral model, post-task perception measures, and regression screens to interpret the institutional mechanism; demographic and country-profile analyses are reported in SI Section~S6. Sample characteristics, missing-data information, and robustness filters are reported in Methods and SI Section~S3; hypothesis-by-hypothesis robustness checks appear in SI Section~S4; and the expanded exploratory analyses appear in SI Sections~S4.3.3, S5, and S6.
\subsection{Democratic redistribution stabilized cooperation, but costly participation narrowed the electorate}
\subsubsection{Cooperation, institutional returns, and democratic participation}
\begin{figure}[p]
\centering
\includegraphics[width=\textwidth]{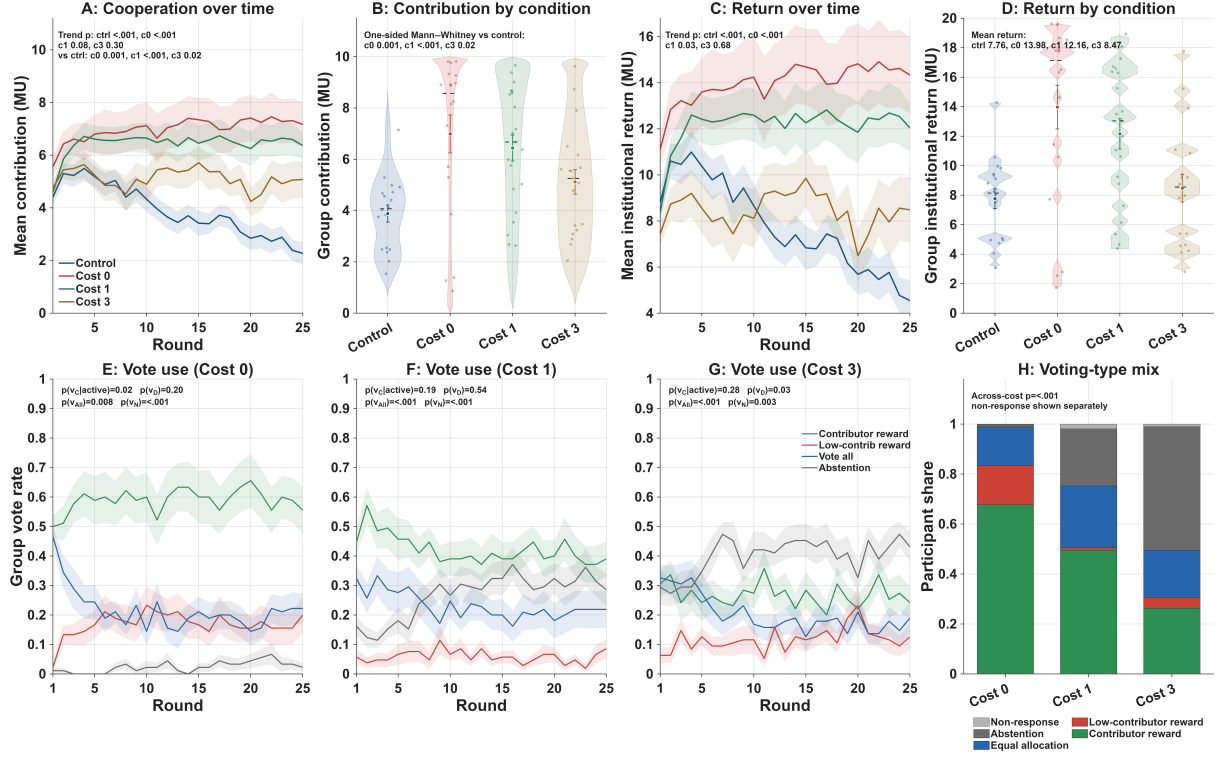}
\caption{Cooperation, participation, and institutional change across conditions. Fig.~\ref{fig:confirmatory}(A) shows mean contribution by round for the no-democracy control and the three democratic cost conditions; shaded bands are standard errors across five-person groups. Fig.~\ref{fig:confirmatory}(B) shows group-level mean contribution distributions with transparent violins, individual group points, mean markers with error bars, and median bars; inset statistics report the preregistered control-versus-democracy comparisons. Fig.~\ref{fig:confirmatory}(C) shows round-level mean institutional return, again with standard-error bands. Fig.~\ref{fig:confirmatory}(D) shows group-level mean institutional-return distributions by condition in the same format as panel B. Fig.~\ref{fig:confirmatory}(E-G) trace the mean group-level rates of contributor-rewarding voting, low-contributor-rewarding voting, vote-to-all, and explicit abstention in each democratic treatment. Fig.~\ref{fig:confirmatory}(H) shows dominant recorded voting-type shares by democratic condition, with explicit abstention and non-response separated. Full definitions, trend tests, pairwise strategy contrasts, and robustness analyses are reported in Methods and SI Section~S4; the preregistered H4 collapse diagnostic is shown in SI Fig.~S22.}
\label{fig:confirmatory}
\end{figure}

The simple public-goods control showed the classic declining pattern \cite{Ledyard,FischbacherGaechter2010}, with contributions falling steadily over rounds (round-level linear trend test of contribution on round, $p<0.001$). Endogenous democratic redistribution changed that trajectory. In the zero-cost democratic condition, contributions increased over time ($p<0.001$). Cost-1 and cost-3 democracies no longer showed the same upward trajectory but neither reproduced the control's steep decay (cost-1, $p=0.080$; cost-3, $p=0.300$; Fig.~\ref{fig:confirmatory}(A-B)). 

Average contributions were also higher under democratic redistribution. Mean contribution was \MeanControlContributionVone{} MUs in the no-democracy control, \MeanCostZeroContributionVone{} MUs at cost 0, \MeanCostOneContributionVone{} MUs at cost 1, and \MeanCostThreeContributionVone{} MUs at cost 3. The group-level mean differences relative to the control were 3.11 MUs for cost 0 (cluster bootstrap 95\% CI, 1.50 to 4.57), 2.56 MUs for cost 1 (1.38 to 3.70), and 1.24 MUs for cost 3 (0.13 to 2.36). Each democratic treatment exceeded the control in the preregistered one-sided Mann--Whitney tests (cost-0, $p=0.001$; cost-1, $p<0.001$; cost-3, $p=0.019$), and that result survived several robustness filters, including the last-10-round window (Fig.~\ref{fig:confirmatory}(B); SI Section~S4; SI Tables~S40 and S41).

Institutional return, defined as the allocated share of the doubled public good minus any voting cost, also declined sharply in the control (round-level linear trend test, $p<0.001$), rose in zero-cost democracy ($p<0.001$), rose more modestly in cost-1 democracy ($p=0.033$), and flattened in cost-3 democracy ($p=0.679$; Fig.~\ref{fig:confirmatory}(C)). Mean institutional return was 7.76 MUs in the control, 13.98 MUs in zero-cost democracy, 12.16 MUs at cost 1, and 8.47 MUs at cost 3. Thus zero-cost and cost-1 democracy increased the return generated by the public-good allocation relative to the control, whereas the high-cost institution increased cooperation but produced a much more modest and statistically uncertain return advantage (cluster bootstrap difference for cost 3, 0.70 MUs, 95\% CI, $-1.44$ to 2.91). We therefore separate three outcomes throughout the paper, namely cooperation, active democratic participation, and institutional return.

Fig.~\ref{fig:confirmatory}(E-G) shows the temporal trajectory of democratic use. Zero-cost democracy stayed heavily contributor-rewarding and kept abstention near zero, whereas both costly democracies showed a strong rise in non-participation. Mean recorded abstention rates rose from 2.4\% of individual voting opportunities at cost 0 to 26.7\% at cost 1 and 40.0\% at cost 3; recorded non-response was much lower, at 0.2\%, 2.2\%, and 1.3\%, respectively (SI Section~S4.1). In cost-1 and cost-3 democracy, the contributor-rewarding share among active votes did not collapse outright. The electorate itself thinned as more participants stopped paying to take part, with round-level abstention trend tests giving $p<0.001$ for cost 1 and $p=0.003$ for cost 3 (SI Table~S43). Institutionally, the cost of democracy mattered not because it immediately turned active voters toward low-contributor-rewarding allocation, but because it gradually encouraged withdrawal.

Across cooperation, institutional return, and participation, the patterns qualitatively matched the weaker-selection graded-contribution regime in Fig.~\ref{fig:evolutionary-selection-comparison}, where voting cost mainly thins participation rather than improving cooperation by filtering active use that rewards low contributors.

This shift is reflected in the composition of dominant voting types (Fig.~\ref{fig:confirmatory}(H)). In Fig.~\ref{fig:confirmatory}(H), each participant is assigned to the democratic action they used most often, either voting to allocate to above-or-equal-to-average contributors, voting to allocate to below-or-equal-to-average contributors, voting to allocate to all, explicitly abstaining, or not recording a voting response. Ties were rare, and the tie rule is reported in SI Section~S5.3; continuous vote-share analyses show the same cost pattern without forcing participants into a single category. Contributor-rewarding voters dominated cost-0 democracy, whereas abstainers became progressively more common as voting grew more expensive. Across conditions, a chi-square test of dominant voting type by cost showed that the strategy distribution changed strongly with cost ($p<0.001$; SI Tables~S35--S37). The same sorting appears in behavior, with contributor-rewarding voters contributing the most, low-contributor-rewarding voters the least, and the high-cost treatment containing many more abstainers than the zero-cost condition (Fig.~\ref{fig:learning}(A); SI Tables~S35--S37). 

The institutional-return profile sharpens this interpretation (Fig.~\ref{fig:learning}(B)). Contributor-rewarding voters were not merely the most cooperative group; they also earned the highest mean institutional return within every democratic cost condition (cost 0, 17.07 MUs; cost 1, 14.57 MUs; cost 3, 11.54 MUs), and this return differed across dominant voting types in all three conditions (Kruskal--Wallis tests, cost 0 and cost 1, $p<0.001$; cost 3, $p=0.0058$; SI Tables~S35 and S36). The contrast was sharpest at high cost. Contributor-rewarding voters still received the highest average return (11.54 MUs), whereas equal-allocation voters and low-contributor-rewarding voters received much less (6.42 and 7.44 MUs, respectively), even though abstainers had become the most common type (7.77 MUs; SI Table~S35). Participation cost therefore did not just reduce the quantity of democratic use. It changed who kept using the institution and what kind of institution democracy became over time.

\subsubsection{Democratic free riding problem and learning dynamics}
These results pose a puzzle. If contributor-rewarding voters earned the highest institutional returns, why did costly democracy produce so much abstention? The answer lies in the fact that this between-participant profile can obscure within-group return incentives and round-to-round learning. We therefore examine those dynamics next.

\begin{figure}[p]
\centering
\includegraphics[width=\textwidth]{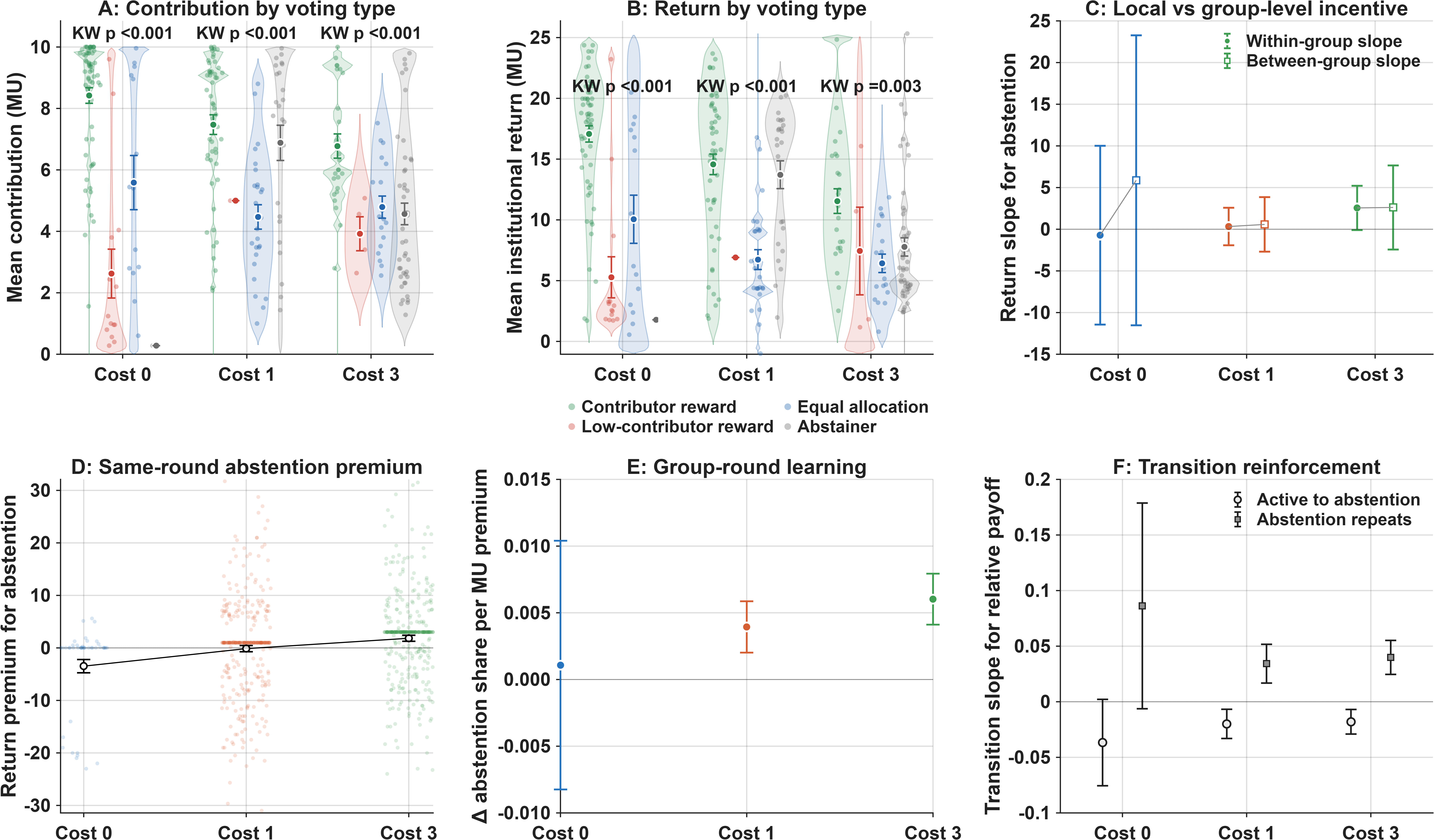}
\caption{Institutional-return incentives and learning behind democratic withdrawal. Fig.~\ref{fig:learning}(A-B) show participant-level mean contribution and mean institutional return by dominant voting type and cost condition, where dominant voting type is each participant's most frequent democratic action, either above-or-equal-to-average allocation, below-or-equal-to-average allocation, equal allocation, or abstention. Violin densities are clipped to the bounded or observed plotting range, and rare cells are shown as points rather than informative distributions. Fig.~\ref{fig:learning}(C) decomposes the association between abstention rate and participant institutional return into a within-group component (abstaining more or less than one's group-members) and a between-group component (belonging to a group with higher mean abstention), with mean contribution controlled; markers show OLS slopes and 95\% confidence intervals. Fig.~\ref{fig:learning}(D) shows, for mixed group-rounds containing both active voters and abstainers, the institutional-return premium for abstention, defined as the mean return of abstainers minus the mean return of active voters. Small points are group-round observations; large markers show condition means with standard errors. Fig.~\ref{fig:learning}(E) tests whether a larger abstention premium in round $t$ predicted an increase in group abstention share in round $t+1$, controlling for current abstention share, round, and group fixed effects. Fig.~\ref{fig:learning}(F) shows participant-level logistic transition slopes for whether relative return predicted switching to abstention in the next round among active voters and abstaining again among abstainers. Exact estimates and $p$ values are reported in SI Tables~S49 and S50.}
\label{fig:learning}
\end{figure}

We separated within-group return incentives from between-group composition (Fig.~\ref{fig:learning}(C-F); SI Table~S49). With explicit abstention separated from non-response, the participant-level within-group slope in the high-cost condition remained positive but imprecise ($\beta=2.56$, $p=0.063$), whereas the corresponding between-group abstention term was not reliable ($\beta=2.61$, $p=0.313$; Fig.~\ref{fig:learning}(C)). This is the democratic free-riding opportunity in behavioral form. Inside a costly institution, abstention can improve the institutional-return component of payoff even while democratic cooperation weakens. The round-level evidence made this local incentive clearer: non-response was rare relative to explicit abstention, same-round high-cost abstention premiums persisted for both institutional return and full round earnings, and higher abstention premiums predicted later withdrawal (SI Section~S4.1; SI Table~S50).

Round-to-round dynamics show how that incentive could be learned. In group-rounds containing both active voters and explicit abstainers, abstention was disadvantageous when voting was free (mean abstainer minus active-voter institutional return $=-3.48$ MUs, one-sample $t$-test, $p=0.009$), neutral at cost 1 ($-0.14$ MUs, $p=0.814$), and advantageous at cost 3 ($1.84$ MUs, $p=0.001$; Fig.~\ref{fig:learning}(D)). The same comparison using full round earnings, defined as the 10-MU endowment minus contribution plus institutional return, weakened the zero-cost disadvantage ($-1.53$ MUs, $p=0.117$), remained non-significant at cost 1 ($0.67$ MUs, $p=0.201$), and remained positive at cost 3 ($2.18$ MUs, $p<0.001$; SI Table~S50). Costly-condition group-rounds in which abstainers received higher returns were then followed by larger increases in abstention in the next round (OLS with group fixed effects, cost 1, $\beta=0.00394$, $p<0.001$; cost 3, $\beta=0.00603$, $p<0.001$; Fig.~\ref{fig:learning}(E)). Individual transitions gave the same reinforcement signal. In cost 1 and cost 3, active voters with lower relative return were more likely to abstain next round, while abstainers with higher relative return were more likely to abstain again (logistic regressions, all $p\le0.003$; Fig.~\ref{fig:learning}(F)). These round-level and transition models are reduced-form exploratory summaries of dependence-rich repeated behavior, so we use them to identify mechanism rather than as separate confirmatory tests.

\subsubsection{An exploratory behavioral model}
\begin{figure}[p]
\centering
\includegraphics[width=\textwidth]{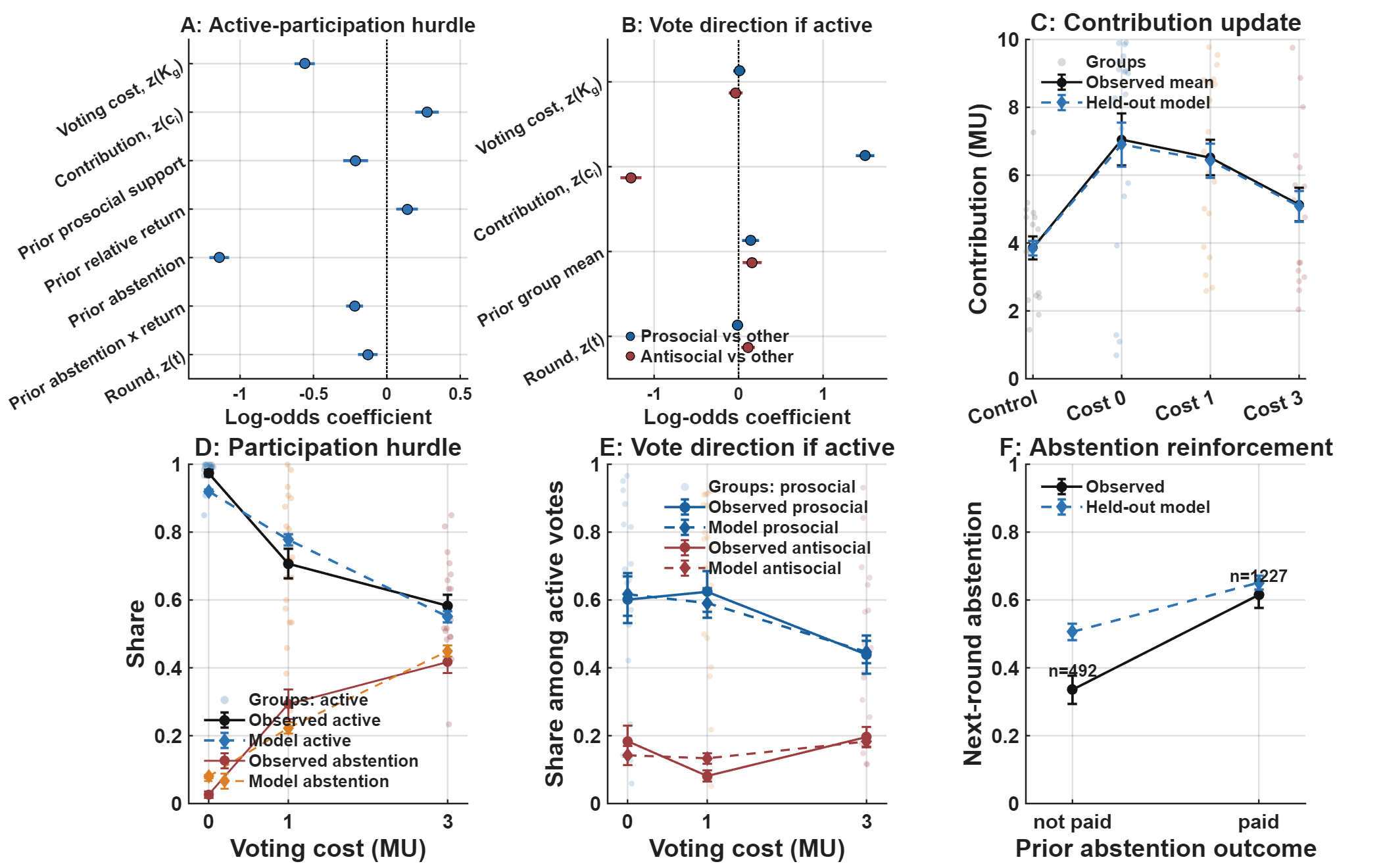}
\caption{Exploratory behavioral model. The model decomposes repeated democratic behavior into contribution updating, active participation, and active-vote direction among those who participate. Fig.~\ref{fig:behavioral-bridge}(A-B) shows the active-participation and active-vote-direction contrasts. Figs.~\ref{fig:behavioral-bridge}(C-F) compare observed group-level summaries with held-out calibrated summaries from five-fold group-level cross-validation for contribution, active voting and abstention, active-vote direction, and next-round abstention after neutral or below-average versus above-average relative return. Small points show group summaries where applicable; markers and error bars show means and standard errors across groups. The model is exploratory and links the evolutionary mechanism to observed experimental behavior. Full equations, coefficient definitions, and calibration checks are reported in Methods and SI Section~S4.3.3.}
\label{fig:behavioral-bridge}
\end{figure}

The analyses above show the behavioral ingredients of democratic free riding, namely conditional cooperation, costly participation, and reinforcement of abstention when abstention pays. To connect those ingredients more directly to the human experiment, we fitted an exploratory behavioral model (Methods; SI Section~S4.3.3) that uses the experimental data to separate three linked behavioral components, covering contribution updating, active participation versus abstention, and vote direction among active voters. Fig.~\ref{fig:behavioral-bridge}(A-B) shows the fitted participation and vote-direction components; Fig.~\ref{fig:behavioral-bridge}(C-F) tests whether the same fitted structure reproduces held-out experimental summaries.

In the active-participation component, voting cost reduced active participation ($\beta=-0.560$ per one-standard-deviation increase in cost, group-cluster bootstrap 95\% CI, -0.745 to -0.414). Prior abstention itself strongly predicted continued withdrawal ($\beta=-1.142$, 95\% CI, -1.323 to -0.931), and the prior abstention-return term remained negative after that persistence term was included ($\beta=-0.220$, 95\% CI, -0.302 to -0.146; Fig.~\ref{fig:behavioral-bridge}(A)). By contrast, in the vote-direction component estimated only among active voters, voting cost did not increase low-contributor-rewarding voting (low-contributor-rewarding-versus-other active vote, $\beta=-0.034$, 95\% CI, -0.303 to 0.229; contributor-rewarding-versus-other active vote, $\beta=0.009$, 95\% CI, -0.235 to 0.222; Fig.~\ref{fig:behavioral-bridge}(B)).

Using the five-fold group-level cross-validation procedure defined in Methods, the same model reproduced the central held-out summaries. Democratic redistribution maintained higher contribution than the control (Fig.~\ref{fig:behavioral-bridge}(C)), active voting declined as voting became costly while abstention rose (Fig.~\ref{fig:behavioral-bridge}(D)), active voters remained more often contributor-rewarding than low-contributor-rewarding (Fig.~\ref{fig:behavioral-bridge}(E)), and prior abstention followed by above-average return was followed by more next-round abstention than prior abstention followed by neutral or below-average return (Fig.~\ref{fig:behavioral-bridge}(F)). The behavioral model therefore links the human data back to the evolutionary results. It converges with the weaker-selection graded-contribution regime by locating the cost effect at participation and abstention reinforcement, and it diverges from the stronger-selection filtering regime because cost did not produce a low-contributor-rewarding shift among those who still voted.

\subsubsection{Dynamic diagnostics}

We tested two dynamic signatures motivated by the evolutionary model, oscillatory instability and bistable institutional states. Evidence for oscillatory movement was limited and is best read as local and group-specific rather than as a condition-wide oscillatory regime (SI Fig.~S23(A-C), SI Fig.~S24, and SI Section~S4). This is not a mismatch with the model because the oscillatory-orbit regimes are idealized mean-field signatures, and experiment-parallel finite-population simulations do not predict synchronized condition-wide oscillations in small populations (SI Section~S1.6).

The bistability analysis was more consistent with the theory. Zero-cost democracy showed the clearest separation into high and low contributor-rewarding late-round states, whereas cost-1 and cost-3 democracy did not show comparably robust two-state structure (SI Fig.~S23(D) and SI Table~S71). In the experimental data, the dynamic signature that most closely matches the model is bistability under zero-cost democracy; costly democracy looked less like a synchronized oscillatory takeover and more like heterogeneous drift toward abstention.

\subsubsection{Preregistered evidence}
The preregistration focused on whether democratic participation sustains cooperation, whether participation cost creates democratic free riding and institutional instability, and whether any intermediate cost is beneficial. The full confirmatory tests and robustness analyses are reported in SI Section~S4 and SI Fig.~S22. Table~\ref{tab:preregistered-map} summarizes the evidence hierarchy. In brief, democracy increased cooperation relative to the no-democracy control, and higher contributors participated more in democratic decision-making. By contrast, the predicted intermediate-cost optimum was not supported. The evidence for democratic free riding was also more nuanced than the preregistered participant-level return comparison suggested because costly voting created a local institutional-return incentive for abstention, even though high-abstention groups did not outperform other groups. Taken together, the preregistered results show that democratic allocation of public-good return promoted cooperation, whereas participation costs mainly narrowed and reshaped the electorate.

\subsection{Costly democracy changed retrospective participation perceptions more than stated norms}

We then asked how participants reconstructed their own behavior and social environment after the task \cite{FischbacherGaechter2010,BicchieriXiao2009,KrupkaWeber2013}. These survey reports should not be read as direct evidence for one psychological process such as memory distortion, self-enhancement, motivated recall, rounding, or norm compliance. They are most useful as institutional traces because they show whether a democratic environment leaves people with accurate or distorted post-task perceptions of how participation was sustained. Participants' reconstructions were informative overall, with reported contributions, vote counts, and stated contribution norms tracking the behavioral record rather than collapsing toward a common midpoint (SI Section~S5). The key question is where those reconstructions systematically departed from the record. We therefore focus on signed deltas, defined as reported minus recorded behavior. Positive values mean that participants reported more than the behavioral record showed. For contribution, $\Delta_{own}$ is reported own contribution minus actual own contribution, and $\Delta_{others}$ is reported average contribution of the other four group members minus their actual average contribution. For voting, $\Delta_{vC}$, $\Delta_{vD}$, $\Delta_{vAll}$, and $\Delta_{vN}$ are reported minus actual counts of contributor-rewarding votes, low-contributor-rewarding votes, equal-allocation votes, and abstentions. Finally, $\gamma$ is the stated appropriate contribution minus actual own contribution. Full definitions and robustness analyses are presented in SI Section~S5.

\subsubsection{Cost-dependent perception gaps}
\begin{figure}[p]
\centering
\includegraphics[width=\textwidth]{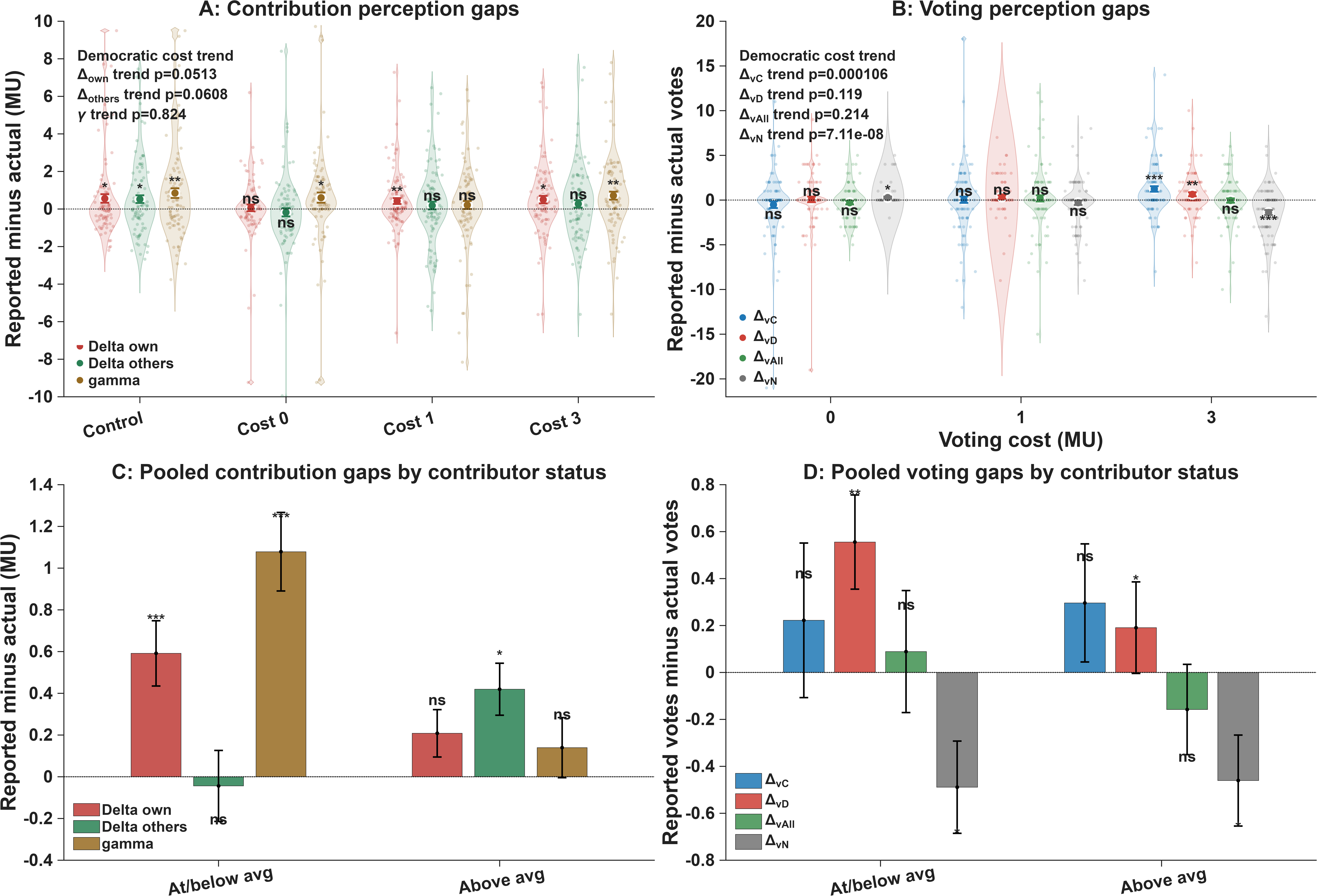}
\caption{Perception gaps and contributor-status differences. Fig.~\ref{fig:delta}(A) shows reported own contribution minus actual contribution ($\Delta_{own}$), reported others' contribution minus the actual mean contribution of the other group members ($\Delta_{others}$), and the gap between stated appropriate contribution and actual contribution ($\gamma$). Fig.~\ref{fig:delta}(B) applies the same signed-gap logic to reported minus actual vote counts for contributor-rewarding votes, low-contributor-rewarding votes, equal-allocation votes, and abstention in the democratic treatments. In panels A and B, violins and small points show participant-level signed gaps, and larger markers with error bars show condition means and standard errors. Fig.~\ref{fig:delta}(C-D) pool democratic participants and compare those who contributed above the average of the other members of their own group with everyone at or below that benchmark. Positive values indicate over-reporting relative to recorded behavior, except for $\gamma$, where positive values indicate a stated norm above recorded own contribution. Stars mark one-sample tests against zero, and printed trend labels report the cost-gradient tests used in the text. The condition-specific contributor-status extension is reported in SI Fig.~S26. Exact tests, robustness checks, and full tables appear in SI Section~S5.}
\label{fig:delta}
\end{figure}

\textbf{Costly voting increased reported own contribution relative to the behavioral record.} Fig.~\ref{fig:delta}(A) and Table~\ref{tab:perception-map} first show what participants reported about contribution. In the no-democracy control, participants significantly over-reported both their own contribution ($\Delta_{own}=\DeltaOwnMeanControlVone{}$, one-sample $t$-test of deviation from zero, $p=0.015$; here and below, small one-sample $p$ values indicate evidence for a non-zero mean signed gap, whereas large values indicate that the mean gap is not statistically distinguishable from zero) and the contribution of the other group members ($\Delta_{others}=\DeltaOthersMeanControlVone{}$, $p=0.013$). In zero-cost democracy, by contrast, both contribution-report gaps were centered near zero ($\Delta_{own}=\DeltaOwnMeanCostZeroVone{}$, one-sample $t$-test, $p=0.735$; $\Delta_{others}=\DeltaOthersMeanCostZeroVone{}$, $p=0.420$). Thus, the large behavioral increase in cooperation under zero-cost democracy did not require participants to inflate either their own contribution or the contribution of others. Once voting became costly, the two contribution-report gaps separated. Participants significantly over-reported their own contribution in cost-1 democracy ($\Delta_{own}=\DeltaOwnMeanCostOneVone{}$, one-sample $t$-test, $p=0.008$) and cost-3 democracy ($\Delta_{own}=\DeltaOwnMeanCostThreeVone{}$, $p=0.014$), whereas reported others' contribution remained statistically centered around the recorded behavior in both costly conditions (cost 1, $\Delta_{others}=\DeltaOthersMeanCostOneVone{}$, $p=0.367$; cost 3, $\Delta_{others}=\DeltaOthersMeanCostThreeVone{}$, $p=0.175$). The democratic cost trend was suggestive but just above the conventional threshold for both $\Delta_{own}$ (Spearman trend test, $p=0.051$) and $\Delta_{others}$ ($p=0.061$), so the clearest contribution-report statement is self-focused over-reporting under costly voting rather than a robust cost-gradient in perceived group cooperation (SI Section~S5; SI Tables~S73 and S74).

\textbf{Stated cooperative norms were institution-sensitive, but cost did not raise them further.} The contribution level participants said was appropriate differed across conditions when tested at the group level (Kruskal--Wallis test on group means, $p=\NormGroupKWPVone{}$). Stated norms were lowest in the control (\NormControlGroupMeanVone{} MUs), highest in zero-cost democracy (\NormCostZeroGroupMeanVone{} MUs), and then declined across the costly democratic conditions (cost 1, \NormCostOneGroupMeanVone{} MUs; cost 3, \NormCostThreeGroupMeanVone{} MUs; democratic cost trend, Spearman $\rho=\NormDemocraticSpearmanRhoVone{}$, $p=\NormDemocraticSpearmanPVone{}$). The task therefore affected stated cooperative ideals, but higher voting cost did not raise those ideals further. The contribution-norm gap $\gamma$ asks the related but distinct question of whether the contribution participants said was appropriate exceeded what they actually contributed. $\gamma$ was significantly positive in the control ($\gamma=\GammaMeanControlVone{}$, one-sample $t$-test, $p=0.001$), zero-cost democracy ($\gamma=\GammaMeanCostZeroVone{}$, $p=0.029$), and high-cost democracy ($\gamma=\GammaMeanCostThreeVone{}$, $p=0.001$), but not in cost-1 democracy ($\gamma=\GammaMeanCostOneVone{}$, $p=0.325$). $\gamma$ did not increase with voting cost (Spearman trend test across democratic costs, $p=0.824$). Costly democracy therefore changed stated norms less directly than it changed retrospective reports of democratic participation (Fig.~\ref{fig:delta}(A); Table~\ref{tab:perception-map}; SI Section~S5).

\textbf{Costly democracy was reported as more active and contributor-rewarding than it was.} Voting reports changed more sharply. Fig.~\ref{fig:delta}(B) and Table~\ref{tab:perception-map} show that the contributor-rewarding-vote report gap $\Delta_{vC}$ increased with voting cost (Spearman trend test, $p<0.001$), while the abstention report gap $\Delta_{vN}$ became more negative (Spearman trend test, $p<0.001$). At cost 3, participants significantly over-reported contributor-rewarding votes (one-sample $t$-test, $p<0.001$) and significantly under-reported abstentions ($p<0.001$). Equal-allocation voting did not show the same cost-sensitive pattern (Spearman trend test, $p=0.214$), and low-contributor-rewarding vote reports were less consistent (Spearman trend test, $p=0.119$). Because voting reports are compositional, we also checked the option-specific gaps against combined reporting-error diagnostics, which preserved the same cost-sensitive interpretation (SI Section~S5; SI Table~S86). These patterns suggest that, when democratic participation became costly, participants increasingly reconstructed themselves as having taken part, and especially as having taken part in contributor-rewarding allocation, more often than the behavioral record showed. The main signed-gap results were also supported by group-cluster bootstrap intervals, including the cost-3 contributor-rewarding-vote gap and cost-3 abstention gap. This main cost-sensitive pattern survived both recency checks and the more detailed use-compression diagnostics (SI Section~S5; SI Fig.~S25; SI Tables~S73, S74, S75, S79, and S86).

\textbf{Robustness checks showed limited recency and self-favoring bias.} We then checked whether participants showed recency or self-favoring bias in their reconstructions. Recent behavior did not account for the democratic participation gaps. When behavioral benchmarks were restricted to the last 10 rounds, some contribution-report gaps shifted, as expected in conditions where contributions changed over time, but the cost-graded voting pattern persisted. Costly democracy still increased the contributor-rewarding-vote report gap and made the abstention report gap more negative. Nor did the contribution-report results reduce to a broad self-favoring account, since $\Delta_{own}-\Delta_{others}$ was close to zero within each democracy and showed no cost gradient. Full analyses are reported in SI Section~S5, SI Fig.~S25, and SI Tables~S78 and S79.

\subsubsection{Higher and lower contributors reconstructed democracy differently}
To distinguish local contributor roles within groups, we classified democratic participants according to whether their mean contribution was above the average contribution of the other members of their own group or at or below that benchmark. This decomposition showed that higher and lower local contributors not only behaved differently, but also reconstructed the institution differently. Above-group contributors contributed more, earned more, and used active votes more often for contributor-rewarding allocation, whereas at-or-below contributors had the larger norm shortfall $\gamma$ (rank-sum $p<0.001$) and above-group contributors had the larger others-report gap $\Delta_{others}$ (rank-sum $p=0.025$; Fig.~\ref{fig:delta}(C-D); SI Table~S94; SI Fig.~S26; SI Tables~S77 and S83). Lower contributors mainly fell short of the contribution level they themselves endorsed, whereas higher contributors tended to report the surrounding group as more cooperative than the behavioral record implied. By contrast, contributor status did not reliably separate voting-perception gaps (SI Table~S94; SI Section~S5.2). The substantive message is that democratic cooperation is partly supported by accurate institutional perception. When participants over-report how much they themselves participated in governance, the institution can look more civically sustained in post-task perceptions than it was in fact.

Contributor status did not reliably explain the voting-perception gaps. Those gaps were organized more by institutional cost (Fig.~\ref{fig:delta}(A-B)) and, secondarily, by dominant voting style than by whether a participant contributed above or at or below the local group average (SI Table~S94; SI Sections~S5 and S5.3).

\subsubsection{Why participants voted}
We next asked why participants used democracy in different ways. SI Table~S95 summarizes the main exploratory answers, and SI Section~S5 reports the full behavior-only, survey-only, and joint regressions. The clearest answer was behavioral rather than dispositional. Participants who cooperated more also used the institution more often for contributor-rewarding allocation.

Round-1 contribution predicted later mean contribution in the no-democracy control and in all three democratic treatments (all $p\le0.023$, all $q\le0.046$; SI Tables~S4--S5). Democracy therefore did not create cooperation from a blank slate. Instead, it changed how initial cooperative differences were amplified, protected, or weakened over repeated interaction.

That continuity helps explain who used the institution to reward contribution. Higher mean contribution predicted more contributor-rewarding votes in all three democratic cost conditions (all $p<0.001$; SI Table~S95; SI Tables~S1 and S3). In zero-cost democracy, higher contributors also cast fewer low-contributor-rewarding votes ($p<0.001$), whereas participants with a larger contribution-norm shortfall $\gamma$ abstained more ($p=0.003$, $q=0.019$; SI Tables~S1 and S13). The same norm gap predicted lower mean contribution in the control, zero-cost democracy, and cost-1 democracy ($p\le0.011$, $q\le0.036$; SI Tables~S4--S6). This behavioral pattern is consistent with a narrower survey signal in zero-cost democracy, where higher SVO predicted higher contribution and fewer low-contributor-rewarding votes (SI Table~S95; SI Tables~S2, S13, and S14). Free democracy thus aligned cooperation and governance by separating a cooperative core that both contributed and used the institution to protect cooperation from participants who endorsed a more cooperative rule than they themselves met.

Therefore, contribution and voting were tightly coupled when participation was easy. However, voting cost progressively uncoupled democratic governance from cooperation by making abstention attractive and by shifting retrospective perceptions of participation. In pooled democratic regressions, higher voting cost predicted fewer contributor-rewarding votes, more abstention, lower contribution, lower institutional return, larger contributor-rewarding-vote report gaps, and stronger under-reporting of abstention (all $p<0.001$; SI Table~S95; SI Figs.~S29--S32; SI Tables~S8--S14).

Notably, equal allocation was not simply another civic option. Higher contributors used vote-to-all less often in every democratic condition (cost 0, $p=0.013$; cost 1, $p<0.001$, $q<0.001$; cost 3, $p=0.002$, $q=0.016$), and in the cost-1 full model vote-to-all predicted lower contribution ($\beta=-0.38$, $p<0.001$, $q=0.002$; SI Table~S6). The democratic options therefore carried distinct behavioral meanings; they were not interchangeable expressions of civic participation.

\subsection{Demographic heterogeneity was local, not the main mechanism}
Demographic variation added texture, but it did not overturn the institutional mechanism. The cooperation benefit of democratic redistribution was visible across the largest participant subsets, and no demographic-by-cost interaction replaced the voting-cost mechanism (SI Section~S6). A behavior-only principal-component and clustering analysis of country profiles provided a broader view: after accounting for voting cost, residence countries differed mainly along a contributor-rewarding cooperation versus equal-allocation axis, with abstention forming a second axis (SI Fig.~S43; SI Tables~S92 and S91). Among countries with larger matched democratic samples, the United Kingdom, Portugal, Poland, and South Africa fell on the more contributor-rewarding cooperative side of this map. Yet the most consistent pattern across countries was the same institutional mechanism seen in the pooled data: as participation became costly, democratic cooperation weakened primarily through withdrawal from active voting (SI Fig.~S44; SI Table~S28).

\section{Discussion}
Endogenous democratic allocation of public-good return creates a double challenge. It can align individual incentives with the public good only if enough contributors continue to use the institution that produces that alignment. We asked whether such an institution can become self-sustaining, or whether the same democratic machinery can instead stabilize an antisocial allocation order or democratic free riding. The evolutionary model showed that all three outcomes are possible. Democratic allocation can support a prosocial order, can be captured by low-contribution strategies, or can allow individuals to benefit from a cooperative allocation rule while avoiding the cost of participation. Voting cost is therefore not mechanically good or bad. Under stronger selection it can select against use of the institution to reward low contributors, whereas under weaker selection, where even free democracy already evolves a more contributor-rewarding electorate, thinning participation reduces contributor-rewarding democratic participation and cooperation.

The human experiment placed behavior within this theoretical map. When participation was free, groups used democratic allocation of public-good return to build a prosocial order. Contributions did not merely stay above the equal-share control; they increased over time, reversing the usual decay of cooperation in repeated public-goods settings. This establishes the constructive force of the institution: endogenous collective choice can transform a vulnerable public good into a cooperative order that directs returns toward those who sustain it \cite{DalBoFosterPutterman,SutterInstitution,KameiDemocracy}.

When voting became costly, democracy did not mainly deteriorate because active voters turned toward low-contributor-rewarding allocation. It deteriorated because participation thinned. Active voting fell, abstention rose, and the cooperative advantage weakened as fewer participants carried the contributor-rewarding institution. This is why democratic free riding appeared most clearly as a local institutional-return opportunity and abstention dynamic rather than as a simple permanent type of high-return nonvoter. The empirical pattern is therefore closer to the weaker-selection participation-thinning regime than to the stronger-selection filtering regime. The relevant failure mode was not an abrupt low-contributor-rewarding capture of the vote, but a gradual narrowing of the electorate that sustained cooperation.

The behavioral model gives this theoretical interpretation a dynamical mechanism. In held-out groups, it reproduced the main treatment summaries and located the cost effect primarily at the active-participation hurdle. Locally profitable abstention predicted later withdrawal, while voting cost did not produce a comparable shift toward low-contributor-rewarding allocation among participants who still voted. Democratic free riding is therefore not only a static payoff category. It can become a dynamic process in which abstention is reinforced when the institution continues to deliver benefits without requiring active support.

Notably, the institution also changed what participants reported about democratic participation. Costly democracy increased the gap between recorded behavior and post-task reports: participants reported more active and contributor-rewarding participation, and less abstention, than the behavioral record showed. The pattern does not identify a single memory mechanism, but it reveals a diagnostic problem. Democratic erosion through withdrawal can be partly invisible to participants themselves if the institution is remembered, or reported, as more actively supported than it was. More broadly, unequal voice, information costs, and participation costs can make the active public less representative than formal democratic rules suggest, so an institution can remain open in principle while the coalition maintaining it becomes smaller and less visible \cite{Lijphart1997,Verba1995,MettlerSoss2004,SchlozmanVerbaBrady2012,Prior2007,KarpowitzMendelbergShaker2012}.

The design also has limits. The treatment is democratic redistribution in a stylized public-goods environment, not democracy in its full political form. Because the experiment was designed to study endogenous democratic redistribution rather than to isolate legitimacy from material incentives, our causal claim is not that voting alone increased cooperation. It is that a democratically chosen redistribution institution increased cooperation relative to a standard equal-share public-goods control, while participation cost narrowed the active coalition that sustained it. The cost manipulation also changed the initial task-currency balance, because costly-voting participants started with enough MU credit to cover the maximum possible voting cost. Because starting balances differed across cost conditions, we cannot fully separate the effect of per-vote cost from the way those costs were framed as deductions from an initial task-currency balance. At the same time, this design makes the withdrawal result conservative in one respect: costly-voting participants always had enough task currency to vote in every round, so abstention cannot be explained by inability to pay. Participation declined even though participation remained feasible throughout the experiment. The voting cost is also a simple monetary cost, whereas real participation costs include information, time, attention, deliberation, and coordination. These limits narrow the causal claim but not the central mechanism, since withdrawal can become locally attractive when maintaining a cooperative institution is individually costly, even while the institution remains collectively valuable.

These findings speak to a familiar challenge in collective governance. Institutions are often designed with good incentives in mind but poor participation economics. Modern democratic systems ask citizens to absorb information, monitor institutions, and act despite limited time and attention. Such costs can make civic voice uneven even when formal rights are equal \cite{Lijphart1997,Verba1995,MettlerSoss2004}. Our results suggest that low-friction democratic participation can be strongly cooperative, whereas even moderate participation cost can hollow out the coalition that sustains the institution. The lesson is not that democratic participation must be effortless, nor that laboratory voting maps directly onto national politics. It is that participation costs are not neutral background details, since they can reshape who governs, how cooperation is sustained, and how people later describe their own democratic engagement. That conclusion should matter wherever collective welfare depends on repeated participation in shared decision-making, from local public goods to online communities and workplace governance \cite{Ostrom,Dietz,DannenbergGallier,Olken2010}.

\section{Methods}
\subsection{Theoretical model}

\subsubsection{Finite-population evolutionary update}
The evolutionary models use a fitness-proportional birth process with mutation. In a finite well-mixed population, individuals are randomly assigned to groups, play the democratic public-goods game, receive payoff, and generate the next generation by parent sampling with probability proportional to fitness. Offspring inherit the parental contribution and voting strategy subject to mutation. The binary version uses eight joint strategies: individuals either contribute ($C$) or defect ($D$), and choose among four voting actions, namely vote for contributors ($vC$), vote for defectors ($vD$), vote for all ($vAll$), or abstain ($vN$). In this binary institution, vote-to-all contributes approval support to both contributor-rewarding and low-contributor-rewarding threshold allocations. The implemented rule is therefore governed by whether contributor-rewarding support exceeds low-contributor-rewarding support, with equal allocation occurring when the two supports tie. This is not the exact three-option plurality rule used in the experiment, where equal allocation is a separate active option and all-abstention triggers a random fallback. The binary model therefore functions as a deliberately distinct approval/threshold institution rather than merely as a simplification. It tests whether the cooperation--participation mechanism persists under a different democratic-allocation rule. The graded-contribution model below then implements the experimental plurality rule, random active-vote tie breaking, and random all-abstention fallback; the analysis in SI Section~S1.6 shows similar cost-sensitive cooperation and participation patterns, suggesting that the mechanism is robust with respect to the details of the allocation rule.

Fitness is an exponential transform of payoff, $w_i=\exp(\beta\pi_i)$, so $\beta$ controls how sharply payoff differences affect reproduction. We use this individual-based process as the finite-population implementation of the same selection--mutation logic captured by replicator--mutation dynamics \cite{TaylorJonker1978,SchusterSigmund1983,HofbauerSigmund1998}. In SI Sections~S1.5 and S1.6, we analyze the finite-population behavior of both the binary strategy set and the graded-contribution extension. In the main text, we focus first on the corresponding infinite-population formulation because it makes the mechanism, attractors, and cost derivatives transparent. The exponential payoff-to-fitness map also lets us vary selection strength explicitly. To check that the mechanism does not depend on this map, we also analyze a payoff-proportional binary update in which parent sampling is directly proportional to payoff rather than to $\exp(\beta\pi)$. This variation preserves the central qualitative patterns, including bistability, oscillatory-orbit regimes, and parameter regions with an intermediate-cost cooperation optimum (SI Section~S1.2).

\subsubsection{Binary infinite-population replicator dynamics}
We analyze the infinite-population limit of the binary individual-based model by tracking the frequencies of the eight joint contribution--voting strategies. Payoffs are expectations over the possible co-player compositions in a well-mixed group. Let $\mathcal{S}$ be the strategy set, let $\mathbf{n}=(n_k)_{k\in\mathcal{S}}$ count the $g-1$ co-players of each strategy, and let $|\mathbf{n}|=g-1$. At population state $\rho$, the probability of composition $\mathbf{n}$ is
\begin{align}
\Pr(\mathbf{n}\mid\rho)=\frac{(g-1)!}{\prod_{k\in\mathcal{S}} n_k!}\prod_{k\in\mathcal{S}}\rho_k^{n_k}.
\end{align}
For a focal individual with strategy $s$, define $N_C(\mathbf{n},s)=z_s+\sum_k n_kz_k$ and $N_D(\mathbf{n},s)=g-N_C(\mathbf{n},s)$. The public good is $G(\mathbf{n},s)=rcN_C(\mathbf{n},s)$. Let $q^C_s=1$ if strategy $s$ votes for contributors or all, and $q^D_s=1$ if strategy $s$ votes for defectors or all. Thus vote-to-all contributes approval support to both threshold allocations. The effective support for contributor allocation is $M_C(\mathbf{n},s)=q^C_s+\sum_k n_kq^C_k$, and the effective support for defector allocation is $M_D(\mathbf{n},s)=q^D_s+\sum_k n_kq^D_k$. The binary institution allocates to contributors when $M_C>M_D$, allocates to defectors when $M_D>M_C$, and splits equally when the two support totals tie. The realized allocation return to the focal individual is
\begin{align}
R_s(\mathbf{n})=
\begin{cases}
G(\mathbf{n},s)/N_C(\mathbf{n},s), & M_C(\mathbf{n},s)>M_D(\mathbf{n},s)\ \hbox{and}\ z_s=1,\\
G(\mathbf{n},s)/N_D(\mathbf{n},s), & M_D(\mathbf{n},s)>M_C(\mathbf{n},s)\ \hbox{and}\ z_s=0,\\
G(\mathbf{n},s)/g, & M_C(\mathbf{n},s)=M_D(\mathbf{n},s),\\
0, & \hbox{otherwise.}
\end{cases}
\end{align}
The expected allocation payoff is $\Omega_s(\rho)=\sum_{|\mathbf{n}|=g-1}\Pr(\mathbf{n}\mid\rho)R_s(\mathbf{n})$. The payoff of strategy $s$ is then
\begin{align}
\pi_s(\rho)=-cz_s-c_v a_s+\Omega_s(\rho),
\end{align}
where $z_s=1$ for contributing strategies and $z_s=0$ for defecting strategies, and $a_s=1$ for active voting and $a_s=0$ for abstention. A common background payoff is not included because it multiplies every exponential fitness by the same factor and cancels from the update. In the infinite-population limit, the frequency of strategy $x\alpha$ after selection and mutation is
\begin{align}
\rho_{x\alpha}(t+1)=\sum_{y,\delta}\nu_{x\alpha}^{y\delta}\rho_{y\delta}(t)\frac{\exp[\beta\pi_{y\delta}(t)]}{\sum_{y,\delta}\rho_{y\delta}(t)\exp[\beta\pi_{y\delta}(t)]}.
\end{align}
Here $x$ and $y$ denote contribution strategies, $\alpha$ and $\delta$ denote voting strategies, $\rho_{x\alpha}$ is the frequency of strategy $x\alpha$, and $\nu_{x\alpha}^{y\delta}$ is the probability that a parent with strategy $y\delta$ produces an offspring with strategy $x\alpha$. The binary phase diagrams use group size $g=5$, contribution cost $c=1$, mutation rate $10^{-3}$ for contribution and voting strategy, and $T=6000$ evolutionary update steps, with stationary averages computed over the last 1200 steps. Phase labels are assigned from the stationary-window cooperator density, cooperator-density amplitude, and spread across initial conditions, so multistable regions are separated from oscillatory-orbit regions. Additional analyses of selection strength and finite-population counterparts are reported in SI Sections~S1.4 and S1.5; the continuous-time exponential-fitness formulation is reported in SI Section~S1.5.1.

We use the same formulation to derive the immediate effect of voting cost. At a fixed population state, write the payoff of strategy $s$ as $\pi_s=\pi_s^0-c_v a_s$, where $a_s=1$ for active voting and $a_s=0$ for abstention. All active votes pay the same cost, so voting cost directly penalizes active voting relative to abstention, not low-contributor-rewarding voting relative to contributor-rewarding voting. For an otherwise identical active voter and abstainer, the direct fitness penalty is $f_{\mathrm{active}}/f_{\mathrm{abstain}}=\exp(-\beta c_v)$. Let $P_s$ denote the post-selection frequency of strategy $s$ before mutation, let $z_s=1$ for contributing strategies and $z_s=0$ for defecting strategies, and let $C^+=\sum_s z_sP_s$ be post-selection cooperation. Differentiating $C^+$ with respect to voting cost gives
\begin{align}
\frac{\partial C^+}{\partial c_v}=-\beta\,\mathrm{Cov}_P(z,a),
\end{align}
where the covariance is taken across strategies using the frequencies $P_s$. A marginal voting cost therefore increases cooperation only when the active strategies being penalized are disproportionately defecting strategies. In both infinite-population replicator dynamics and finite-population well-mixed simulations, this occurs most clearly in the stronger-selection regime. That regime therefore acts as a filtering regime, in which a small voting cost can increase cooperation and contributor-rewarding use of democracy. Selection strength enters Eq.~(5) through $\beta$: stronger selection turns the same contribution--participation association into a sharper change in post-selection strategy frequencies, whereas weaker selection makes the response to cost more gradual. We numerically validate this derivative and the corresponding finite-population behavior in SI Sections~S1.3, S1.4, and S1.5.

\subsubsection{Graded-contribution finite-population model}
The graded-contribution model checks whether the same mechanism survives a richer action space matching the experiment. Each individual has an integer contribution strategy $h_i\in\{0,\ldots,10\}$ and a voting strategy $a_i\in\{v_C,v_D,v_{All},v_N\}$. In each generation, a well-mixed population of size $N$ is randomly partitioned into groups of five. The active option with the most votes wins; ties among active options are broken at random; if all group members abstain, one of the three active allocation rules is selected at random. Evolutionary payoff is full round payoff, $\pi_i=10-h_i-c_v\mathbf{1}\{a_i\neq v_N\}+R_i$, where $R_i$ is the allocated public-good return. Fitness is $w_i=\exp(\beta\pi_i)$. The next generation is formed by sampling parents with probability proportional to $w_i$. Contribution mutation moves the contribution strategy to an adjacent integer level with reflecting boundaries at 0 and 10, and vote mutation switches the voting strategy to one of the other three voting actions.

The main graded-contribution comparison uses $N=1000$, $T=5000$ generations, and the last 1000 generations as the stationary window. We chose the focal parameter set through a simulation search over selection strength, mutation rates, and initial strategy distributions, targeting the qualitative ordering of contribution, active voting, abstention, and contributor-rewarding support across costs rather than a point-by-point numerical fit to the experiment (SI Section~S1.6). For the weaker-selection comparison in Fig.~\ref{fig:evolutionary-selection-comparison}, we use $\beta=0.008$, contribution mutation $\mu_C=0.120$, vote mutation $\mu_V=0.003$, initial contributions $h_i\sim\mathrm{Binomial}(10,0.85)$, and initial vote probabilities $(0.70,0.10,0.15,0.05)$ for $(v_C,v_D,v_{All},v_N)$. The stronger-selection contrast keeps the same mutation and initialization structure but sets $\beta=0.10$, so payoff differences have a larger effect on parent sampling. SI Section~S1.6 also reports graded-contribution initial-condition screens over the full $r-c_v$ plane, population-size screens, and example stationary time series.

\subsection{Experimental procedure}
The experiment received ethical approval from the Ethics Committee of the University of Konstanz (IRB statement 29/2023). The preregistered experiment \cite{Salahshour2024} was implemented as an oTree app \cite{ChenSchongerWickens2016} and run online between 13 and 21 June 2024. Participants were recruited through Prolific \cite{PalanSchitter2018}, provided informed consent before participation, and assigned to a no-democracy control or one of three democratic voting-cost conditions of 0, 1, or 3 MUs.

In each round, participants received 10 MUs and chose a contribution between 0 and 10. Contributions were doubled. In the no-democracy control, the multiplied public good was divided equally among all group members. In the democratic conditions, participants chose one of four voting options: allocate the public good equally among participants whose contribution was above or equal to the group average, allocate it equally among participants whose contribution was below or equal to the group average, allocate it equally among all group members, or abstain. Participants exactly at the group average were included in either threshold category if that category won. The option with the most active votes determined the allocation; ties among active options were broken at random. If every group member abstained, one of the three active allocation rules was implemented at random. This all-abstention fallback was rarely invoked, occurring in 0.8\% of democratic group-rounds overall, with 0.0\%, 0.2\%, and 2.1\% in cost 0, cost 1, and cost 3, respectively (SI Section~S4). Abstention was an explicit voting option. We distinguish explicit abstention (recorded vote 4) from non-response or no recorded voting response (recorded vote 0) in the voting-composition robustness checks; the primary participation measures treat both as non-participation, and Fig.~\ref{fig:confirmatory}(H) displays them separately. Voting options 1--3 carried the cost in the costly treatments; abstention did not. After each round, participants received feedback on contribution, vote, income, and the relevant group outcomes.

Participants started the experiment with a balance equal to the maximum possible voting cost over the 25 rounds, giving 0 MUs in the control and cost-0 conditions, 25 MUs in the cost-1 condition, and 75 MUs in the cost-3 condition. The balance was task-currency credit used to cover possible voting costs. Participant account balances and payment reconciliation are documented separately in SI Section~S3. The task instructions stated that accumulated MUs were converted into a GBP bonus. Each MU was paid at 1/288 GBP, approximately GBP 0.00347. Thus, voting cost was manipulated in the task currency but the marginal GBP value of one MU was not confounded with condition. The experiment lasted a median of 32.8 min in matched Prolific records. The participation fee was approximately GBP 3.28, with individual values depending on completion time. The median variable bonus was GBP 1.60 (mean GBP 1.61; range GBP 0.24--2.71).

\subsection{Post-experiment surveys}
The task-specific survey asked participants to estimate their own average contribution, other players' average contribution, appropriate contribution, and, in democratic conditions, perceived vote success, the most appropriate vote, the income-maximizing vote, and how often they used each voting option. Social preferences were measured with the Social Value Orientation slider task \cite{MurphySVO}. The subsequent questionnaire contained an 11-item conformity/susceptibility block based on Mehrabian and Stefl \cite{MehrabianStefl} and an adapted 15-item horizontal/vertical individualism--collectivism block based on Singelis \emph{et al.} and Triandis and Gelfand \cite{SingelisHVIC,TriandisGelfand}. Risk preference was elicited with an incentivized five-step adaptive choice between a 50\% lottery and a sure payment following the logic of Holt and Laury \cite{HoltLaury}. The survey sources are detailed in SI Section~S7.

\subsection{Data filtering and statistical analysis}
The primary confirmatory sample comprised \NIncludedParticipantsVone{} participants organized into \NIncludedGroupsVone{} complete five-person groups. The condition-specific sample sizes were 18 groups in the no-democracy control, 18 in cost-0 democracy, 21 in cost-1 democracy, and 19 in cost-3 democracy. Four included groups contained a participant inactive for at least 10 contribution rounds; we used this inactivity threshold as a robustness filter reported in SI Section~S4.1 and SI Table~S34. Detailed sample counts and missing-data summaries are reported in SI Section~S3 and SI Table~S30. Prolific demographic information was linked by participant ID and anonymized before analysis. Demographic information was available for \NDemoMatchedVone{}/\NDemoTotalVone{} behavioral participants, and age was reported by 365 participants (mean \DemoAgeMeanVone{} years, SD \DemoAgeSdVone{}). Sex was recorded as male for 187 participants, female for 175, and prefer not to say for 3.

Hypothesis 1 used one-sided Mann--Whitney tests comparing group-level mean contribution in each democratic condition with the control; the one-sided direction was preregistered. Hypothesis 2 used one-sided Mann--Whitney tests comparing low-frequency and high-frequency voters in the costly conditions, with supplementary round-level voter versus non-voter institutional-return tests and participant-level return regressions. Hypothesis 3 used Spearman correlations between contribution and democratic participation. Hypothesis 4 used a chi-square test relating high non-participation to preregistered collapse status, with extended collapse analyses under global and condition-specific thresholds. Hypothesis 5 used the preregistered intermediate-cost contrast plus planned concavity checks. Hypothesis 6 used one-way ANOVA across democratic cost conditions at the preregistered group level, complemented by analyses of contributor-rewarding share among active votes and active participation. The main confirmatory comparisons use the five-person group as the statistical unit. Participant-level, group-round, and round-level mechanism analyses are labeled exploratory and use dependence-aware specifications where possible, including group fixed effects, group-level summaries, or clustered interpretation of repeated observations. Exploratory perception analyses used signed deltas, one-sample tests against zero, cost-trend tests, between-group comparisons, and group-cluster bootstrap intervals for the main signed gaps \cite{EfronTibshirani1993}.

Exploratory regression analyses used standardized predictors so that effects measured in different units could be compared on a common scale. We used Benjamini--Hochberg false-discovery-rate correction within related analysis families, which limits the expected share of false positives among discoveries while retaining power for exploratory tests \cite{BenjaminiHochberg1995}. The main families were tests tied to the preregistered hypotheses, perception-gap analyses, behavioral-model coefficients, demographic heterogeneity analyses, and country-profile analyses. Full details are reported in SI Sections~S4, S4.3.3, S5, S5, and S6.

\subsection{Exploratory behavioral model}
The evolutionary models identify three ingredients of democratic cooperation: contribution updating, the decision to keep voting when voting is costly, and the direction chosen by those who remain active. The experiment contains the same ingredients in a finite repeated interaction. We therefore used an exploratory behavioral model to describe how these ingredients appeared in observed round-to-round choices. The model is a descriptive bridge between the evolutionary mechanism and the experiment, not a preregistered hypothesis test.

For participant $i$ in group $g$ and round $t$, let $c_{igt}$ denote contribution, let $A_{igt}$ be an indicator that equals 1 for an active vote and 0 for abstention, and let $Y_{igt}\in\{C,D,All\}$ denote the active vote direction. Here $C$ denotes a contributor-rewarding vote, $D$ a low-contributor-rewarding vote, and $All$ equal allocation. Let $K_g$ denote voting cost, $S_{g,t-1}$ previous contributor-rewarding support among active voters in the group, $N_{ig,t-1}=1$ previous abstention, $\Delta\pi_{ig,t-1}$ previous institutional return relative to the other members of the group, $R_{g,t-1}$ the previous implemented allocation rule, $\bar c_{-i,g,t-1}$ the previous mean contribution of the other group members, and $\bar c_{g,t-1}$ the previous group mean contribution. We write probabilities on the log-odds scale, $\ell(x)=\log[x/(1-x)]$, because this is equivalent to a sigmoid choice rule: each probability is generated by a latent behavioral index, and additive changes in history or incentives shift that index while the resulting probability remains between 0 and 1 \cite{Luce1959,McFadden1974,Train2009}. Continuous predictors are standardized as $z(x)=(x-\bar x)/s_x$ within the estimation sample, where $\bar x$ and $s_x$ are the sample mean and sample standard deviation of $x$, so coefficients compare one-standard-deviation changes in variables measured on different scales.

The first component of the model describes contribution updating. Because contributions were integer Money Units from 0 to 10, we represent them as ten unit-scale contribution opportunities with probability $p_{igt}$:
\[
\begin{aligned}
c_{igt} &\sim \mathrm{Binomial}(10,p_{igt}),\\
\ell(p_{igt})&=\alpha+\phi z(c_{ig,t-1})+\lambda z(\bar c_{-i,g,t-1})+\theta_{R_{g,t-1}}+\delta_{\mathrm{condition}}+\delta_t z(t).
\end{aligned}
\]
This component allows contribution to depend on a participant's own previous contribution, the previous contribution of the other group members, the previously implemented allocation rule, condition, and time.

The second component describes the participation hurdle: whether the participant casts an active vote rather than abstaining,
\[
\begin{aligned}
A_{igt} &\sim \mathrm{Bernoulli}(q_{igt}),\\
\ell(q_{igt})&=\beta+\gamma_{K_g}+\gamma_c z(c_{igt})+\gamma_s z(S_{g,t-1})+\gamma_\pi z(\Delta\pi_{ig,t-1})\\
&\quad+\gamma_N z(N_{ig,t-1})+\gamma_{N\pi} z(N_{ig,t-1}\Delta\pi_{ig,t-1})+\gamma_t z(t).
\end{aligned}
\]
This component separates the decision to participate in governance from the direction of the vote. The interaction $N_{ig,t-1}\Delta\pi_{ig,t-1}$ asks whether abstention becomes self-reinforcing when abstention previously paid relative to the rest of the group.

The third component describes vote direction among active voters. Define $I^C_{igt}=\mathbf{1}\{Y_{igt}=C\}$ and $I^D_{igt}=\mathbf{1}\{Y_{igt}=D\}$. Conditional on active voting, the two vote-direction contrasts are
\[
\begin{aligned}
\Pr(I^C_{igt}=1\mid A_{igt}=1)&=m^C_{igt},\\
\ell(m^C_{igt})&=\eta_C+\omega_C z(K_g)+\rho_C z(c_{igt})+\zeta_C z(\bar c_{g,t-1})+\tau_C z(t),\\
\Pr(I^D_{igt}=1\mid A_{igt}=1)&=m^D_{igt},\\
\ell(m^D_{igt})&=\eta_D+\omega_D z(K_g)+\rho_D z(c_{igt})+\zeta_D z(\bar c_{g,t-1})+\tau_D z(t).
\end{aligned}
\]
In the contributor-rewarding contrast, $I^C_{igt}=0$ includes low-contributor-rewarding and equal-allocation active votes. In the low-contributor-rewarding contrast, $I^D_{igt}=0$ includes contributor-rewarding and equal-allocation active votes. Equal-allocation votes are therefore retained as active votes and serve as part of the comparison category in both contrasts. This lets the model ask whether voting cost changes active participation itself, or instead changes the direction chosen by those who remain active.

For held-out calibration, we partitioned whole five-person groups into five folds \cite{Stone1974}. All participants and all rounds from a group were held out together. This is stricter than holding out individual observations because choices within a group are interdependent; the calibration therefore asks whether the model reproduces behavior in groups that did not contribute to model fitting. Error bars in Fig.~\ref{fig:behavioral-bridge}(C-F) are standard errors across group-level summaries; key coefficients used group-cluster bootstrap intervals. Full behavioral model definitions and coefficient contrasts are reported in SI Section~S4.3.3.

\section*{Acknowledgements}
This work was supported by funding from the Deutsche Forschungsgemeinschaft (DFG, German Research Foundation) under Germany's Excellence Strategy -- EXC 2117-422037984, the Deutsche Forschungsgemeinschaft Gottfried Wilhelm Leibniz Prize 2022 584/22 (I.D.C.), the Max Planck Society, the European Union's Horizon 2020 Research and Innovation Programme under the Marie Sklodowska-Curie Grant agreement no. 860949, the Struktur- und Innovationsfonds fuer die Forschung of the State of Baden-Wuerttemberg, the PathFinder European Innovation Council Work Programme no. 101098722, the Office of Naval Research Grant N0001419-1-2556, and Small Project Grant (S23-32) from the Center for the Advanced Study of Collective Behavior.

\clearpage

\clearpage

\begin{table}[!htbp]
\centering
\caption{Preregistered hypotheses and support status. Directional tests were preregistered; exploratory extensions are reported as mechanism checks rather than confirmatory replacements.}
\label{tab:preregistered-map}
\small
\setlength{\tabcolsep}{3pt}
\renewcommand{\arraystretch}{0.95}
\begin{tabularx}{0.995\textwidth}{@{}>{\raggedright\arraybackslash}p{0.8cm}>{\raggedright\arraybackslash}p{3.9cm}>{\raggedright\arraybackslash}p{2.15cm}>{\raggedright\arraybackslash}X@{}}
\toprule
Hyp. & Prediction & Status & Main interpretation\\
\midrule
H1 & Democratic decision-making increases cooperation relative to no democracy. & Supported & All three democratic conditions produced higher group-level cooperation than the control.\\
H2 & Costly democracy gives democratic low-frequency voters higher institutional returns than frequent voters. & Partly supported & The preregistered participant-level contrast was weak, but within-group and round-to-round analyses showed a local institutional-return incentive for abstention when voting was costly.\\
H3 & Higher contributors participate more in democratic decision-making. & Supported & Higher contributors used active votes more for contributor-rewarding allocation and abstained less, especially before high cost made participation fragile.\\
H4 & Democratic free riding predicts institutional collapse. & Partly supported & Collapse-style diagnostics were informative but heterogeneous; high-cost dynamics looked more like drift toward abstention than a synchronized collapse.\\
H5 & Intermediate voting cost optimizes cooperation by reducing free riding. & Not supported & Zero-cost democracy produced the strongest cooperation pattern.\\
H6 & Voting cost moderates contributor-rewarding democratic participation. & Partly supported & Cost changed participation mainly by thinning the active electorate, not by simply converting active voters into low-contributor-rewarding voters.\\
\bottomrule
\end{tabularx}
\end{table}

\begin{table}[!htbp]
\centering
\caption{Perception--behavior gaps by institution. Each cell reports the condition mean of a signed gap and the one-sample $t$-test of whether that mean differs from zero. For these one-sample entries, small $p$ values indicate evidence that the condition mean deviates from zero, whereas large values indicate no statistically detectable deviation from zero. Positive values mean that participants reported more than the recorded value, except for $\gamma$, where positive values mean that the participant's stated appropriate contribution exceeded their actual contribution. Democratic trend $p$ values are Spearman tests across voting costs 0, 1, and 3. The recency row is a contribution-report sensitivity check in which cells report the change in $\Delta_{own}$ when the behavioral benchmark is the last 10 rounds rather than all 25 rounds, with paired-test $p$ values; the democratic column reports the pooled democratic paired test. Voting-count recency checks are reported in SI Fig.~S25 and SI Table~S79. Main-text $p$ values are rounded to three decimals, with smaller values reported as $p<0.001$. Green and red denote significant positive and negative deviations at $p<.05$; amber denotes $0.05\le p<.10$; grey denotes non-significant deviations.}
\label{tab:perception-map}
\scriptsize
\resizebox{\textwidth}{!}{%
\begin{tabular}{p{2.8cm}p{2.3cm}p{2.3cm}p{2.3cm}p{2.3cm}p{2.1cm}p{4.7cm}}
\toprule
Short name & Control & Cost 0 & Cost 1 & Cost 3 & Democratic trend / pooled test & Main reading\\
\midrule
Own-report gap ($\Delta_{own}$) & \textcolor[rgb]{0.10,0.45,0.18}{+\DeltaOwnMeanControlVone{}; $p=0.015$} & \textcolor[rgb]{0.45,0.45,0.45}{+\DeltaOwnMeanCostZeroVone{}; $p=0.735$} & \textcolor[rgb]{0.10,0.45,0.18}{+\DeltaOwnMeanCostOneVone{}; $p=0.008$} & \textcolor[rgb]{0.10,0.45,0.18}{+\DeltaOwnMeanCostThreeVone{}; $p=0.014$} & \textcolor[rgb]{0.80,0.45,0.05}{$p=0.051$} & Own contribution was accurately centered in zero-cost democracy, but reported as higher than recorded once voting became costly.\\
Others-report gap ($\Delta_{others}$) & \textcolor[rgb]{0.10,0.45,0.18}{+\DeltaOthersMeanControlVone{}; $p=0.013$} & \textcolor[rgb]{0.45,0.45,0.45}{\DeltaOthersMeanCostZeroVone{}; $p=0.420$} & \textcolor[rgb]{0.45,0.45,0.45}{+\DeltaOthersMeanCostOneVone{}; $p=0.367$} & \textcolor[rgb]{0.45,0.45,0.45}{+\DeltaOthersMeanCostThreeVone{}; $p=0.175$} & \textcolor[rgb]{0.80,0.45,0.05}{$p=0.061$} & Perceived cooperation by others rose only weakly with cost; the self-report effect was clearer.\\
Norm shortfall ($\gamma$) & \textcolor[rgb]{0.10,0.45,0.18}{+\GammaMeanControlVone{}; $p=0.001$} & \textcolor[rgb]{0.10,0.45,0.18}{+\GammaMeanCostZeroVone{}; $p=0.029$} & \textcolor[rgb]{0.45,0.45,0.45}{+\GammaMeanCostOneVone{}; $p=0.325$} & \textcolor[rgb]{0.10,0.45,0.18}{+\GammaMeanCostThreeVone{}; $p=0.001$} & \textcolor[rgb]{0.45,0.45,0.45}{$p=0.824$} & Participants often endorsed a higher contribution than they actually made, but this norm--behavior gap did not grow monotonically with cost.\\
Self--other asymmetry ($\Delta_{own}-\Delta_{others}$) & \textcolor[rgb]{0.45,0.45,0.45}{+\DeltaSelfMinusOthersMeanControlVone{}; $p=0.842$} & \textcolor[rgb]{0.45,0.45,0.45}{+\DeltaSelfMinusOthersMeanCostZeroVone{}; $p=0.205$} & \textcolor[rgb]{0.45,0.45,0.45}{+\DeltaSelfMinusOthersMeanCostOneVone{}; $p=0.250$} & \textcolor[rgb]{0.45,0.45,0.45}{+\DeltaSelfMinusOthersMeanCostThreeVone{}; $p=0.322$} & \textcolor[rgb]{0.45,0.45,0.45}{$p=0.682$} & Democracy produced a mild self-favoring asymmetry, but not a cost-graded one.\\
Recency pull ($\Delta_{own}^{last10}-\Delta_{own}^{all}$) & \textcolor[rgb]{0.10,0.45,0.18}{+0.873; $p<0.001$} & \textcolor[rgb]{0.65,0.15,0.10}{-0.255; $p=0.042$} & \textcolor[rgb]{0.45,0.45,0.45}{-0.098; $p=0.384$} & \textcolor[rgb]{0.45,0.45,0.45}{+0.144; $p=0.246$} & \textcolor[rgb]{0.45,0.45,0.45}{pooled $p=0.327$} & Recent-round benchmarks shift some condition estimates, especially the declining control and zero-cost democracy, but do not create a uniform democratic recency bias.\\
Contributor-rewarding-vote report gap ($\Delta_{vC}$) & -- & \textcolor[rgb]{0.45,0.45,0.45}{\DeltaVCMeanCostZeroVone{}; $p=0.179$} & \textcolor[rgb]{0.45,0.45,0.45}{+\DeltaVCMeanCostOneVone{}; $p=0.914$} & \textcolor[rgb]{0.10,0.45,0.18}{+\DeltaVCMeanCostThreeVone{}; $p<0.001$} & \textcolor[rgb]{0.05,0.35,0.70}{$p<0.001$} & Reported contributor-rewarding voting increased with voting cost.\\
Low-contributor-rewarding-vote report gap ($\Delta_{vD}$) & -- & \textcolor[rgb]{0.45,0.45,0.45}{+\DeltaVDMeanCostZeroVone{}; $p=0.802$} & \textcolor[rgb]{0.80,0.45,0.05}{+\DeltaVDMeanCostOneVone{}; $p=0.063$} & \textcolor[rgb]{0.10,0.45,0.18}{+\DeltaVDMeanCostThreeVone{}; $p=0.006$} & \textcolor[rgb]{0.45,0.45,0.45}{$p=0.119$} & Low-contributor-rewarding voting was also over-reported at high cost, but less cleanly than contributor-rewarding voting.\\
Equal-allocation report gap ($\Delta_{vAll}$) & -- & \textcolor[rgb]{0.45,0.45,0.45}{\DeltaVAllMeanCostZeroVone{}; $p=0.147$} & \textcolor[rgb]{0.45,0.45,0.45}{+\DeltaVAllMeanCostOneVone{}; $p=0.591$} & \textcolor[rgb]{0.45,0.45,0.45}{\DeltaVAllMeanCostThreeVone{}; $p=0.828$} & \textcolor[rgb]{0.45,0.45,0.45}{$p=0.214$} & Equal-allocation voting did not show the same cost-sensitive reporting pattern.\\
Abstention report gap ($\Delta_{vN}$) & -- & \textcolor[rgb]{0.10,0.45,0.18}{+\DeltaVNMeanCostZeroVone{}; $p=0.010$} & \textcolor[rgb]{0.45,0.45,0.45}{\DeltaVNMeanCostOneVone{}; $p=0.218$} & \textcolor[rgb]{0.65,0.15,0.10}{\DeltaVNMeanCostThreeVone{}; $p<0.001$} & \textcolor[rgb]{0.05,0.35,0.70}{$p<0.001$} & High-cost participants reported abstaining less than they actually did.\\
\bottomrule
\end{tabular}}
\end{table}

\end{document}